\documentclass[12pt]{iopart}
\usepackage{amsfonts}
\newcommand{\bee}{\begin{eqnarray}}
\newcommand{\eend}{\end{eqnarray}}
\newcommand{\rmp}{{\rm p}}
\newcommand{\Sch}{Schr${\rm \ddot{o}}$dinger equation}
\newcommand{\BS}{the Bethe-Salpeter equation}
\newcommand{\al}{alling to the center}
\begin{document}
\vspace{1cm} \title[Positronium collapse]
 {Positronium collapse in hypercritical magnetic field and restructuring
 of the vacuum in QED}
\author{A E Shabad*  and  V V Usov\dag
}
\address{* \; P.N.Lebedev Physics
Institute, Moscow, Russia 
}
\address{\dag ~ Weizmann Institute of Science, Rehovot, Israel
 }
 \eads{\mailto{*~shabad@lpi.ru},\\
 \hspace{1.2cm} \mailto{\dag~fnusov@wicc.weizmann.ac.il}}
\begin{abstract}
The Bethe-Salpeter equation in a strong magnetic field is studied
for positronium atom in an ultra-relativistic regime, and a
(hypercritical) value for the magnetic field is determined, which
provides the full compensation of the positronium rest mass by the
binding energy in the maximum symmetry state. The compensation
becomes possible owing to the falling to the center phenomenon.
The relativistic form in two-dimensional Minkowsky space is
derived for the four-dimensional Bethe-Salpeter equation in the
limit of an infinitely strong magnetic field, and used for finding
the above hypercritical value. Once the positronium rest mass is
compensated by the mass defect the energy barrier separating the
electron-positron system from the vacuum disappears. We thus
describe the structure of the vacuum in terms of strongly
localized states of tightly mutually bound (or confined) pairs.
Their delocalization for still higher magnetic field, capable of
screening its further growth, is discussed.
\end{abstract}

\section{Introduction} In the present paper we are considering, in
the framework of quantum electrodynamics, the phenomenon of
falling to the center in a system of two charged particles caused
 by the ultraviolet singularity $1/x^2$ of the photon
propagator - which mediates the interaction between them - on the
light-cone, $x^2=x^2_0-{\bf x}^2\simeq 0$. Here $x_0$ is the time,
and $\bf x$ is the space coordinate. This phenomenon occurs in a
number of problems. In some of cases listed below the matter comes
to the one-dimensional \Sch ~with the singular attractive
potential $U(r)=-\beta/r^2$, $0<r<\infty$, the falling to the
center depending on the value of the constant $\beta$ that fixes
the strength of the coupling \cite{QM}.

Long ago it was established  that the falling to the center is
inherent in \BS ~for electron-positron system treated in a fully
relativistic way and taken in the ladder approximation, provided
that the fine structure constant $\alpha$ exceeds a critical value
$\alpha_{\rm cr}$, $(1/137)\ll\alpha_{\rm cr}<1$. We refer to the
so-called Goldstein solution \cite{goldstein} of the Bessel
differential equation in the variable $s=\sqrt{-x^2},~$ the
space-like interval between the particles, to which the problem is
reduced in the ultra-relativistic limit. F\al~ implies that the
singular attraction is so strong that the energy spectrum becomes
unlimited from below, and hence - once the singular problem is
duely defined - the binding energy may completely cancel the rest
mass, so that the gap between positive and negative continua
disappears. This reminds to some extent the situation in the
problem of an electron in an external electric field, occupying
the whole space, where the tunnelling through the gap leads to the
vacuum instability with respect to $e^+e^--$pair creation known as
Schwinger effect, although the latter does not have a threshold
character. The value $\alpha=\alpha_{\rm cr}$ may delimit the
range of self-consistency of the theory known as quantum
electrodynamics, at least in its customary form.

Another well-known theater \cite{popov}, \cite{greiner},
\cite{shabad2} where the f\al~ acts is the point-like nucleus with
its charge $Z$ exceeding the critical value $Z_{\rm
cr}=\alpha^{-1}=137$, into whose Coulomb field a relativistic
electron described by the Dirac equation is placed. After this
problem is properly regularized, the electron energy level reaches
the border
 $\varepsilon=-m$ of the lower continuum as the charge $Z$ grows,
 which gives rise to
 electron-positron pair production, the free positrons leaving the
 atom. This mechanism leads to diminishing the charge and restores
 it at the critical value that cannot thus be exceeded.

 If an electron in an atom is treated within the \Sch, or a pair of
 particles is addressed in a semi-relativistic way with the aid of
 \BS, wherein the famous Salpeter's equal-time Ansatz is made
 (resulting in the disregard of the retardation effects), the f\al
 ~does not take place. The matter is that the singularity
 in the point $r=0$,
 $r=\sqrt{{\bf x}^2}$
 of the Coulomb potential $U(r)=-\alpha Z/r$ originating in this case from
 the light-cone singularity of the photon propagator, mentioned above, is
 not sufficiently strong, and the energy level remains shallow. On the contrary, in the above two
 cases the f\al~ may be attributed to relativistic enhancement of
 the Coulomb force - the second-order differential equation to
 which the spherically symmetric Dirac equation may be reduced by
 excluding the second spinor component, contains the sufficiently singular
 attractive term  $-(\alpha
 Z/r)^2$ in the effective potential \cite{popov}.

 The situation changes drastically when a strong external magnetic
 field  ${\bf B}$ is applied to the system. Already the \Sch ~for a particle
 situated in the combination of the Coulomb and strong magnetic
 fields possesses the f\al ~caused by the singularity $1/|z|$ in the
 differential \Sch ~defined on the whole axis $-\infty<z<\infty$. Here $z$
 is the electron coordinate along the axis parallel to the
 magnetic field. The energy spectrum of this equation contains
 large negative values tending to minus infinity \cite{loudon}
 as the magnetic field $B\rightarrow\infty$. The reason lies in
 the dimensional reduction: for strong magnetic field the electron
 is restricted to the lowest Landau level, consequently its wave
 function obeys a two-dimensional (one space and one time
 coordinates) differential equation. The reduction of the number of
 degrees of freedom causes the effective strengthening of the
 attraction, like in the quark confinement problem.
  Analogous strengthening takes place when the
 electron-positron system is described by  \BS ~studied using
 the semi-relativistic Ansatz that implies the non-relativistic
 character of the relative motion of the constituent particles
 \cite{leinson1} -
 ~\cite{leinson2}. Again,
 the energy
 spectrum of the corresponding two-dimensional equation is
 unlimited from below as the magnetic field grows. The total rest
 mass of the ground state is compensated  when the magnetic field
 reaches the value
 \cite{ShUs},
 whose order of magnitude is determined by the large factor $\exp
 (\rm{const}/\alpha)$. 
 As pointed in \cite{ShUs} 
 those results, however, are not reliable,
 since they depend on the unrighteous extrapolation of non-relativistic
 procedure to the essentially relativistic region of large mass defect.

 The relativistic treatment of an electron in a combination of
 external Coulomb and magnetic field was given in \cite{semikoz}.
The result was that if the magnetic field is reasonably strong
($\sim 10^{15}$ G) the critical value of the nucleus charge $Z$
may correspond to stable elements with $Z\sim 90$.

In the present paper we consider the system of two charged
relativistic particles - especially the electron and positron - in
interaction with each other, when placed in a strong constant and
homogeneous magnetic field $B$. To this end we use \BS ~ in the
ladder approximation without exploiting any non-relativistic
assumption. We derive its limiting - when $B\rightarrow\infty$ -
form, which comes out to be a Bethe-Salpeter equation in
two-dimensional Minkowsky space-time, covariant under the
corresponding Lorentz subgroup - the boost along the magnetic
field. Stress, that the two-dimensionality holds only with respect
to the degrees of freedom of charged particles, while the photons
remain 4-dimensional in the sense that  the singularity of the
photon propagator is determined by the inverse d'Alambert operator
in the 4-dimensional, and not two-dimensional Minkowsky space.
(Otherwise it would be weaker). The term responsible for
interaction with an arbitrary electric field directed along ${\bf
B}$ is also included and does not lay obstacles to the dimensional
reduction. Throughout the paper we set $\hbar=c=1$ and use the
Heaviside-Lorentz units, where $\alpha=e^2/4\pi.$
$\left.B\right|_{\rm Gaussian}=\sqrt{4\pi} \left.B\right|_{\rm
Heaviside-Lorentz}.$

We make sure that in the case under consideration the critical
value of the coupling constant is zero, $\alpha_{\rm cr}=0$, i.e.,
the f\al ~is present already for its genuine value $\alpha=1/137$,
in contrast to the no-magnetic-field case, where $\alpha_{\rm
cr}>1/137$. If the magnetic field is large, but finite, the
dimensional reduction holds everywhere except a small neighborhood
of the singular point $s=0$, wherein the mutual interaction
between the particles dominates over their interaction with the
magnetic field. The dimensionality of the space-time in this
neighborhood remains to be 4, and its  size is determined by the
Larmour radius $L_B=(eB)^{-1/2}$ that is zero in the limit
$B=\infty$. The latter supplies the singular problem with a
regularizing length. The larger the magnetic field, the smaller
the regularizing length, the deeper the level. We find the value
of the magnetic field - we call it \textit{first hypercritical
field} - \bee\label{final0} B^{(1)}_{\rm
hpcr}=\frac{m^2}{4e}\exp\left\{\frac{\pi^{3/2}}
{\sqrt{\alpha}}+2C_{\rm E}\right\},\quad \eend where $C_{\rm
E}=0.577$ is the Euler constant, that provides disappearance of
the center-of-mass energy of the electron-positron pair and of its
center-of-mass momentum component along $\bf B$. We refer to this
situation as a collapse of positronium.

In discussing the physical consequences of the f\al ~we appeal to
the approach recently developed by one of the present authors as
applied to the \Sch ~with singular potential \cite{shabad} and to
the Dirac equation in supercritical Coulomb field \cite{shabad2}.
Within this approach the singular center looks like a
 black hole. The solutions of the differential
equation that oscillate near the singularity point are treated as
free particles emitted and absorbed by the singularity. This
treatment becomes natural after the differential equation is
written as the generalized eigenvalue problem with respect to the
coupling constant. Its solutions make a Hilbert space and are
subject to orthonormality relations with a singular measure. This
singularity makes it possible for the oscillating solutions to be
 normalized  to  $\delta$-functions, as free particle
wave-functions should be. The nontrivial, singular  measure that
appears in the definition of the scalar product of quantum states
in the Hilbert space of quantum mechanics introduces the geometry
of a black hole of non-gravitational origin and the idea of
horizon. The deviation from the standard quantum theory manifests
itself in this approach only when particles are so close to one
another that the mutual Coulomb field they are subjected to falls
beyond the range, where the standard theory may be referred to as
firmly established \cite{shabad2}.

Within this approach the regularizing  length provided by the
Larmour radius is dealt with not as a cut-off, but as a lower
border of the normalization volume, the event horizon in a way.
Although the result (\ref{final0}) is obtained following the
concept of Refs. \cite{shabad}, \cite{shabad2}, it can be
reproduced without essential alteration within the standard
cut-off philosophy, too.

The most intriguing question is what happens after the magnetic
field exceeds the first hypercritical value (\ref{final0}). The
solution of \BS ~in two-dimensional space-time in the
ultra-relativistic limit studied in the present paper corresponds
to formation  of  special "confined" states in the kinematical
domain called sector III in \cite{shabad}, \cite{shabad2}. (Within
the standard approach these would be bound states, although this
is less adequate). As the corresponding overall energy and
momentum of such $e^+e^-$-state is zero, it is not separated from
the vacuum by an energy barrier. Besides, this state is the one of
maximum symmetry in the coordinate and spin space. Hence, it may
be thought of as relating to the vacuum, as well, and describing
its structure. The confined particles cannot escape to infinite
distance from one another, on the contrary the probability density
of the confined state is concentrated near the point $s=0$, behind
the horizon - as distinct from the ordinary bound state.

The situation is expected to change as the magnetic field goes on
growing. At a certain stage - we reserve the name \textit{second
hypercritical} for the corresponding value of the magnetic field -
deconfinement of the above strongly localized states may occur.
The corresponding solutions to the Bethe-Salpeter equation are not
yet strictly obtained, which makes us describe the deconfinement
more hypothetically. After the level deepens further, the
center-of-mass 2-momentum  gets into sufficiently far space-like
region, and solutions oscillating at large distances between
electron and positron appear. Thus, the state delocalizes. The
delocalized electron-positron pairs produced from the vacuum, each
particle  on a Larmour orbit, should screen the magnetic field and
stop its growing above the second hypercritical value, this value
being the absolute maximum of the magnetic field admitted within
quantum electrodynamics. Simultaneously, the space-like total
momentum provides the lattice structure to the vacuum.

This resembles the case of supercritical nucleus where there are
states (that belong to sector IV in the nomenclature of
Refs.\cite{shabad}, \cite{shabad2}) admitting the leakage to
infinity, which provides the mechanism for reducing the  charge of
the nucleus below the critical value.

 No sooner than the delocalized states are found in our present
problem one may definitely claim the instability of the vacuum
with the second hypercritical magnetic field or - which is the
same - the instability of such
 field under the pair creation that might
provide the mechanism for its diminishing. For the present, we
state that the first hypervalue (\ref{final0}) is such a value of
the magnetic field, the exceeding of which would already cause
restructuring of the vacuum and  demand a profound revision of
quantum electrodynamics.

The paper is organized as follows. In Section 2 we revisit
Goldstein's solution by referring to various possibilities in
approaching the f\al, especially the one invoked by the previous
work \cite{shabad}, \cite{shabad2}.

In Section 3 we derive the ultimate two-dimensional form of \BS
~in its differential version characteristic of the ladder
approximation, when the magnetic field tends to infinity, with the
help of expansion over the complete set of Ritus matrix
eigenfunctions \cite{ritus}. The latter accumulate the spacial and
spinor dependence on the transversal-to-the-field degree of
freedom. The Fourier-Ritus transform of the Bethe-Salpeter
amplitude obeys an infinite chain of coupled differential
equations that decouple in the limit of large $B$, so that we are
left with one closed equation for the amplitude component with the
Landau quantum numbers of the electron and positron both equal to
zero, while the components with other values of Landau quantum
numbers vanish in this limit. The resulting equation is a
differential equation with respect to two variables that are the
differences of the particle coordinates: along the time
$t=x_0^\rme-x_0^\rmp$ and along the magnetic field
$z=x_3^\rme-x_3^\rmp$. It contains only two Dirac matrices
$\gamma_0$ and $\gamma_3$ and can be alternatively written using
$2\times 2$ Pauli matrices. Arbitrary external electric field $\bf
E$ along $\bf B$ is also included, $E\ll B$. By introducing
different masses the resulting two-dimensional equation is easily
modified to cover also the case of an one-electron atom in strong
magnetic field and/or other pairs of charged particles.

In Section 4 the ultra-relativistic  solutions (possessing maximum
symmetry) to the equation derived in Section 3 are depicted
corresponding to the vanishing energy-momentum of the
$e^+e^-$-state, and the first hypercritical magnetic field is
found basing on the standing wave boundary condition imposed on
the lower border of the normalizing volume - as prescribed by the
theory in Refs. \cite{shabad}, \cite{shabad2}. Also the standard
cut-off procedure of Refs. \cite{QM}, \cite{popov} is fulfilled to
give practically the same value (\ref{final0}). Further, we
estimate possible modifications that might be introduced by
radiative corrections to the mass operator, to find that these
cannot change the conclusions any essentially, and discuss the
deconfinement.

\section{Bethe-Salpeter equation
for positronium in ultrarelativistic regime}  The fully
relativistic Bethe-Salpeter equation for a system of two Fermions
of masses $m_{\rm a}$ and $m_{\rm b}$ and opposite charges has in
the ladder approximation the form \cite{schweber}
\bee\label{schweber}\hspace{-3cm}[(\rmi\widehat{\partial}_{\rm
a}-m_{\rm a})]_{\alpha \beta} [(\rmi\widehat{\partial}_{\rm
b}-m_{\rm b})]_{\mu\nu}\begin{tabular}{c}{\Large$\chi$}
$_{\beta\nu}(x_{\rm a},x_{\rm b})$\end{tabular} =-\rmi 8\pi\alpha
D_{m n}(x_{\rm a}-x_{\rm
b})\gamma^m_{\alpha\beta}\gamma^n_{\mu\nu}
\begin{tabular}{c}{\Large$\chi$}$_{\beta\nu}(x_{\rm a},x_{\rm b})$\end{tabular}.
\eend Here $\gamma$'s are the Dirac gamma-matrices, the two-time
wave function {\Large$\chi$}$(x_{\rm a},x_{\rm b})$ is a $4\times
4$ matrix with respect to spinor indices, the corresponding Greek
letters running the values (1,2,3,4). The summation over the
repeated vector indices $m,n=0,1,2,3$ is also meant. The
derivatives\bee\label{derivatives}\widehat{\partial}_{\rm a,
b}=\gamma_0\partial ^0_{\rm a, b}-\gamma_i\partial^i_{\rm a,
b},\quad i=1,2,3\eend act on the first and the second arguments of
{\Large$\chi$}$(x_{\rm a},x_{\rm b})$, respectively, $\alpha$ is
the fine structure constant, and $D_{m n}(x_{\rm a}-x_{\rm b})$ is
the photon propagator. Note that the Feynman photon Green function
$D_{\rm F}$ of Ref.\cite{schweber} is defined as 2$D$. The
translational invariance implies that the solution, which is an
eigenfunction of the translation operator, should have the
form\bee\label{trans}\begin{tabular}{c}{\Large$\chi$}$_P(x_{\rm
a},x_{\rm b})$
\end{tabular}=\exp
\left( \rmi P\frac{x_{\rm a}+x_{\rm b}}{2}\right)~\eta_P(x),\eend
where $x=x_{\rm a}-x_{\rm b}.$

Equation (\ref{schweber}) is a complicated set of differential
equations, which can, however, be essentially simplified, if one
assumes $P^\mu=0$ as explained in the previous subsection. In this
case, equation (\ref{schweber}) can be transcribed as
\bee\label{schweber2}
(\rmi\overrightarrow{\widehat{\partial}}-m_{\rm a})\eta_0(x)
(-\rmi\overleftarrow{\widehat{\partial}}^{\rm T}-m_{\rm b})=-\rmi
8\pi \alpha D_{m n}(x)\gamma^m\eta_0(x)(\gamma^n)^{\rm T}, \eend
where the superscript T means transposition and the derivative
acts on the relative variable $x$ to the right or to the left of
it according to what is prescribed by the direction of the arrow.
With the help of the known relation \cite{greiner} \bee\label{C}
\gamma_n^{\rm T}=-C^{-1}\gamma_nC,  \eend where $C$ is the charge
conjugation matrix, eq.(\ref{schweber2}) is transformed
to\bee\label{schweber3}
(\rmi\overrightarrow{\widehat{\partial}}-m_{\rm a})\eta_0(x)C^{-1}
(\rmi\overleftarrow{\widehat {\partial}}-m_{\rm b})=
\rmi8\pi\alpha D_{m n}(x)\gamma^m\eta_0(x)C^{-1}\gamma^n.\eend The
Bethe-Salpeter amplitude $\eta_0(x)C^{-1}$ is the one for the case
where the two Fermions are Dirac conjugated to one another, i.e.
are electron and positron, equation (\ref{schweber3}) with $m_{\rm
a}=m_{\rm b}=m$ being just the Bethe-Salpeter equation for
electron-positron system \cite{greiner}.

We take the photon propagator in the Feynman gauge
$D_{mn}(x)=g_{mn}D(x^2)$, where the metric tensor obeys diag $
g_{mn}=(1,-1,-1,-1)$. The Ansatz\bee\label{ansatz}
\eta_0(x)=\Theta(x)\gamma_5C,\eend where $\Theta(x)$ is a unit
matrix containing a single function of $x$, is then consistent
with the set (\ref{schweber3}) and turns it into the
equation\bee\label{theta}(-\Box+m^2)\Theta (x)+32\rmi\alpha\pi
D(x^2)\Theta(x)=0,\eend where $\Box =-\partial^2_0+{\Delta}$ is
the Laplace operator. The figure 32 here has resulted from the
multiplication of 8 in (\ref{schweber3}) by 4, which is associated
with the dimension of the space:\bee\label{four}
\sum_{m,n=0,1,2,3} g_{mn}\gamma_m\gamma_n=4\eend

The  photon propagator is singular on the light cone $x^2=0$:
\bee\label{photon} D(x^2)=\frac{-\rmi}{4\pi^2x^2}.\eend In the
space-like region $x^2<0$, for the most symmetrical state where
the solution does not depend on the angles in the 4-dimensional
Euclidean space, $\Theta(x)=\Theta(s)$, $s=\sqrt{-x^2}$, equation
(\ref{theta}) with eq.(\ref{photon}) for $D(x^2)$ becomes the
Bessel differential equation \bee\label{last}
-\frac{\rmd^2\Theta}{\rmd s^2}-\frac 3{s}\frac{\rmd\Theta}{\rmd
s}+m^2\Theta=\frac{8\alpha}{\pi s^2}\Theta.\eend Its solution is
known as Goldstein's solution \cite{goldstein} to the
Bethe-Salpeter equation with $P_\mu=0$. (See the review
\cite{seto}, where other ultrarelativistic solutions,
corresponding to Ans$\ddot{\rm a}$tze different from the above are
also listed). Near the singular point $s=0$ equation (\ref{last})
has the asymptotic form \bee\label{asymptotic}
-\frac{\rmd^2\Theta}{\rmd s^2}-\frac 3{s}\frac{\rmd\Theta}{\rmd
s}=\frac{8\alpha}{\pi s^2}\Theta(s),\eend which is also the
asymptotic form of the full (with $P_\mu\neq 0$) Bethe-Salpeter
equation (\ref{schweber}). Its solutions behave near the singular
point $s=0$ like $s^\sigma$, where \bee\label{sigma}
\sigma=-1\pm\sqrt{1-\frac{8\alpha}{\pi}}\eend The substitution
$\Theta(s)=\Psi(s)s^{-\frac {3}{2}}$ eliminates the first
derivative and reduces equation (\ref{last}) to the standard form,
\bee\label{standard}-\frac{\rmd^2\Psi}{\rmd s^2}-\frac 1{s^2}
\left(\frac{8\alpha}\pi-\frac 3{4}\right){\Psi} =-m^2\Psi,\quad
0<s<\infty,\eend of a
 Schr$\ddot{\rm o}$dinger-like equation with purely centrifugal -
attractive or repulsive, depending on the sign of the difference
$(\frac{8\alpha}{\pi}-\frac 3{4})$ - potential. Its solutions
behave near the singular point $s=0$ like\bee\label{sigma2}
\Psi(s)~\sim ~s^{\frac 1{2}\pm\sqrt{1-\frac{8\alpha}\pi}}.\eend
The same as for the usual radial  Schr$\ddot{\rm o}$dinger
equation, the natural mathematical  requirement that the norm
$\int_0|\Psi(s)|^2\rmd s$ be convergent at the lower limit $s=0$
is not yet sufficient to rule out the more singular solution,
which corresponds to the lower sign in (\ref{sigma2}), in the
whole range where the square root in (\ref{sigma}), (\ref{sigma2})
is real (in the present case this requirement would separate the
less singular solution only for formally negative $\alpha$ !). To
do so an additional physical requirement concerning the behavior
of the wave function near the origin is usually imposed to fix the
eigenvalue problem in quantum mechanics \cite{QM}. For the
Bethe-Salpeter equation such physical requirement was established
by Mandelstam \cite{mandelstam}. It reads that the integral over a
small three-dimensional closed hypersurface $S(3)$ of the 4-vector
current density $\overline{\Theta}(x)\partial_\mu\Theta(x)$ around
the origin
\bee\label{mandelstam}\oint\overline{\Theta}(x)\partial_\mu\Theta(x)\rmd
\sigma_\mu\simeq
s^{\pm2\sqrt{1-\frac{8\alpha}\pi}}\int\rmd\Omega\eend should tend
to zero together with the radius $s$ of the hypersphere $S(3)$ in
the Euclidean 4-space-time. This implies that $\Theta(s)$ should
increase slower than $s^{-1} $ and makes only solutions with the
upper sign acceptable, provided that $\alpha<\alpha_{\rm cr}$,
where \bee\label{alphacr}\alpha_{\rm cr}=\frac\pi{8}.\eend The
above requirement also rules out the both oscillating solutions
when $\alpha\geq\alpha_{\rm cr}$. If, however, one keeps to a
weaker condition that the current flow (\ref{mandelstam}) be
\textit{finite} as $s\rightarrow 0$ and, correspondingly,
$\Theta(s)$ increase no faster than $s^{-1}$, the both oscillating
solutions are compatible with it (note the complex conjugation
sign in (\ref{mandelstam}), due to which the square root does not
appear in it if $\alpha\geq \alpha_{\rm cr}$). Such situation is
typical of the falling to the center phenomenon.

The full Bethe-Salpeter equation (\ref{schweber}) certainly has
bound states, when $\alpha<\alpha_{\rm cr}$, corresponding to
positronium atom, and the above condition serves to define them.
The binding energy of the realistic, $\alpha=1/137$, positronium
makes about a millionth fraction of its rest mass. On the
contrary, there is no bound state described by equation
(\ref{last}). The exact solution to the latter, decreasing at
infinity, is expressed in terms of the McDonald function  as
$~\Theta(s)= (1/s)K_{\sqrt{1-\frac{8\alpha}\pi}}(ms)~.$ Its
asymptotic behavior near $s=0$ is a linear combination of  the
both terms (\ref{sigma}) and, hence this solution is forbidden by
the boundary conditions discussed above, provided that $\alpha
<\alpha_{\rm cr}$. This means that no bound state exists with
$P_0={\bf P}=0$ for the coupling, smaller than the critical value
$\alpha=\alpha_{\rm cr}$ (\ref{alphacr}), and the gap between
electrons and positrons survives down to this value.

Our main concern is about what happens in the overcritical region
of the coupling constant $\alpha\geq \alpha_{\rm cr}$. Three
different approaches have the right to exist for treating this
case.

The first is the same as the one used for considering a Dirac
electron in the Coulomb field of a nucleus with its  charge
greater than $1/\alpha$, $Z>137$ (see \cite{popov},
\cite{greiner}). In that approach the finite size of the nucleus
was exploited as providing a natural cut-off to the singular
potential. There is no such natural fundamental length in our
problem, but if we introduce it formally, for instance by shifting
the pole in the photon propagator (\ref{photon}) from the light
cone inwards the time-like domain, $D_{mn}(x^2)\asymp
1/(x^2-\lambda^2)$ we would come to the situation where the
positronium gradually approaches the  point $P_0={\bf P}=0$ as
$\alpha$ grows and reaches it at certain $\alpha_{\rm
cr}=\alpha(\lambda)$. In other words, where there is falling to
the center, the attraction is so strong that for a sufficient
value of the coupling  the level becomes so deep as to fully
compensate for the whole mass of the positronium atom. When
analogous situation  occurs in the above case of the supercharged
nucleus, the electron level dips into the Dirac sea of positron
states, the vacuum becomes unstable with respect to $e^+e^-$-pair
creation.
 The
essential disadvantage of this approach is that all important
quantities, the pair production probability among them, depend on
the cut-off length and do not have a definite limit when
$\lambda\rightarrow 0$. This fact makes the results doubtful after
the cut-off length becomes less than the characteristic length of
the problem, which is the electron Compton length $m^{-1}$. This
is the case for the (supercritical) nucleus, whose size is adopted
to be $10^{-12}cm\ll m^{-1}$. Down to what other border should one
believe the result once it does not converge?

The second approach might be dependent on the von Neuman technique
of the so called self-adjoint extension of the Hamiltonian (see
the pioneering works \cite{case}, \cite{meetz} and the monograph
\cite{reed}). According to \cite{case} and \cite{meetz} there is
an infinite number of discrete  eigenvalues of the operator, whose
differential expression is determined by the left-hand side of
equation (\ref{standard}). These extend to $m^2=\infty$ for fixed
$\alpha$. The self-adjoint extension is defined up to an arbitrary
parameter that can always be chosen in such a way as to make any
given $-m^2$ - the electron mass squared - an eigenvalue. Then one
should conclude that the tightly bound states exist beyond the
point $\alpha=\pi/8$. The disadvantage of this approach is in that
no physical criterion is known to fix the arbitrariness of the
self-adjoint extension \cite{reed}. If one likes to take the above
prescription seriously, one would have to alter the choice of this
parameter when going to a different $\alpha$ or $m^2$. We do not
know if the method of self-adjoint extension was ever applied to
the problem under consideration. In the problem of supercritical
nucleus the application of this method yields the result that the
electron level never sinks into the continuum \cite{case}, hence
there is no place for the electron-positron pairs production
effect.

A third approach is, in our opinion, most adequate. It is to treat
equation (\ref{standard}) as an eigenvalue problem with respect to
the coupling constant $\alpha$. By putting this eigenvalue problem
in the form of an integral equation it was demonstrated
\cite{tiktopoulos}, \cite{seto} that the corresponding integral
kernel gives rise to a self-adjoint operator, once an appropriate
norm is finite, and the eigenvalues $\alpha<\alpha_{\rm cr}$ make
a discrete set. If extended to the supercritical region
$\alpha>\alpha_{\rm cr}$, this procedure may be thought of to be
equivalent to the approach recently developed by one of the
present authors \cite{shabad} wherein equation (\ref{standard}) is
to be represented - by bringing the singular term $-8(\alpha/\pi
s^2)\Psi$ to the right-hand side, and taking the term $-m^2\Psi$
to the left-hand side -  as the generalized eigenvalue problem for
the differential operator $-\frac{\rmd^2}{\rmd s^2}+\frac
3{4s^2}+m^2$ defining the spectrum of $\alpha$. This operator is
self-adjoint in the (rigged) Hilbert  space of functions,
orthonormalizable \textit{with the singular measure} $s^{-2}\rmd
s$ and subjected to the boundary condition
\bee\label{stand}\Psi(s_0)=0 \eend imposed at the lower edge
$s=s_0$ of a normalization box. As long as $s_0$ is finite we face
a discrete spectrum of $\alpha$. The eigen-solutions are standing
waves at the lower edge of the box and decrease at
$s\rightarrow\infty$ like $\exp (-ms)$. In the limit
$s_0\rightarrow 0$ the levels condense to make a
\textit{continuum} of states in the supercritical region
$\alpha>\alpha_{\rm cr}$. The norm of the state vector calculated
with the  singular measure\footnote{The requirement of square
integrability with the above singular measure, when extended back
to the values $\alpha<\alpha_{\rm cr}~$ ($s_0=0$ in this case),
just excludes the lower sign in (\ref{sigma2}) and makes the
imposing of the conditions (\ref{mandelstam}) discussed above for
the bound state problem unnecessary.} diverges in this limit, what
makes it possible to normalize the solution to $\delta$-function
and hence interpret it as corresponding to a free particle living
mostly near the singularity. This situation refers to the
kinematical domain called sector III in \cite{shabad}. The
corresponding solutions were called confined states, since their
wave-function decreases at infinity like that of bound states.
Near the origin  there are free particles, incoming from the
origin, then scattered elastically inwards and then outgoing back
to the origin.

When applied to the supercritical nucleus \cite{shabad2}, this
approach led to the effect of absorption of electrons by the
nucleus and to the known effect of electron-positron pair
production, the corresponding probabilities being calculated in a
cut-off-independent way. These effects, however, occurred in
another kinematical domain, called sector IV, where the particles
are free also at large distances from the singular center. To see,
if analogous effects are intrinsic to the positronium in the
supercritical case $\alpha>\alpha_{\rm cr}$ and cause an
instability of the vacuum state relative to the pair production it
would be necessary to go beyond the kinematical restriction
$P_\mu=0$, since we need solutions, oscillating at infinity as
well as near zero to this end. Leaving this task for future, we
now consider the features of the confined state.

The solution to equation (\ref{standard}) for $\alpha>\alpha_{\rm
cr}$ that decreases when $s\rightarrow\infty$ is given by the
McDonald function  with imaginary index
\bee\label{mcdonald}\Psi(s)=\sqrt{s}\;K_{\rmi\sqrt{
\frac{8\alpha}{\pi}-1}}(ms).\eend This function is real. The lower
edge of the normalization box should be chosen much smaller than
the characteristic length of the problem, which is the electron
Compton length, $s_0\ll m^{-1}$. Then one can use the asymptotic
form of the McDonald function near zero to write the standing wave
boundary condition (\ref{stand}) as an equation for determining
the spectrum of $\alpha$ \bee\label{spectrum}
\left(\frac{ms_0}{2}\right)^{2\nu}=
\frac{\Gamma(1+\nu)}{\Gamma^*(1-\nu)},\quad
\nu=\rmi\sqrt{\frac{8\alpha}{\pi}-1}\eend or
\bee\label{spectrum2}\nu\ln\frac {ms_0}{2}=\rmi\arg\Gamma(\nu+1)-
\rmi \pi n,\qquad n=0,\pm 1,\pm 2,...\eend Confining ourselves to
the values of the coupling constant that do not differ much from
the critical value, $|\nu|\ll 1$ we may exploit the approximation
for the Euler $\Gamma$-function\bee\label{euler}
\Gamma(1+\nu)\cong 1-\nu C_{\rm E},\eend where $C_{\rm E}=0.577$
is the Euler constant, to
get\bee\label{spectrum3}\ln\left(\frac{ms_0}{2}\right)=\frac{-\pi
n}{\sqrt{\frac{8\alpha_n}{\pi}-1}}-C_{\rm E}, \quad n=1,2,...\eend
We have expelled the non-positive integers $n$ from here, since
they would lead to the roots for $ms_0$ of the order of or much
larger than unity in contradiction to the adopted condition
$s_0\ll m^{-1}$. For such values the asymptotic representation of
the McDonald function used above is not valid. It may be checked
that there are no other zeros of McDonald function, besides those
in (\ref{spectrum3}), enumerated by positive integers. Finally,
the discrete spectrum of the coupling constant above the critical
value $\pi/8$ close to it is given as
\bee\label{spectrum4}\alpha_n=\frac
\pi{8}+\frac{\pi}8\frac{n^2}{(\ln\frac 2{ms_0}-C_{\rm E})^2},\quad
n=1,2,3...\eend It is seen explicitly, that the eigenvalues do
condense when the lower edge of the normalization box $s_0$ tends
to zero. The wave function (\ref{mcdonald}) is mostly concentrated
inside the Compton radius $m^{-1}$, but the probability density is
confined to the region close to $s=s_0\rightarrow 0$ due to the
singularity of the measure $s^{-2}$ near this point.

The solution of the Bethe-Salpeter equation, originally written
for positronium atom, relates in the ultra-relativistic situation
$P_\mu=0$ considered, as a matter of fact to the vacuum as well.
Indeed, the state described has no total energy and no total
momentum and correspondingly does not depend on the coordinate
sums of the constituting particles. It is not separated from the
vacuum by an energy barrier. Besides, it is proportional to the
unit matrix in the spinor space and is O(3.1)-invariant, i.e.
possesses the maximum symmetry, as the vacuum should do.
On the other hand, this state has a nontrivial space-time
structure, described by the dependence on the coordinate
differences, which implies the concentration of the state near
zero value of the relative coordinate. These considerations may
mean  a need of the restructuring of the vacuum when the coupling
constant exceeds the critical value and serve to establish the
band of values for this constant beyond which the existing
standard concepts no longer hold. This is how the things stand
with the positronium atom.

In the next section we derive a two-dimensional analog of equation
(\ref{schweber}) relating to the case where a strong magnetic
field is imposed, and describe all solutions corresponding to the
fall-down to the center. The two-dimensioning makes this
phenomenon stronger. We shall return to the analysis of its
consequences in the subsequent section. The important difference
with the present situation will be that the agent providing the
fall-down to the center will be the external magnetic field,
whereas the coupling constant will be kept equal to its
experimental value $\alpha=1/137$ throughout.

\section{Derivation of two-dimensional Bethe-Salpeter equation in
asymptotically strong magnetic field }

The view that charged particles in a strong constant magnetic
field are confined to the lowest Landau level and behave
effectively as if they possess only one spacial degree of freedom
- the one along the magnetic field - is widely accepted. Moreover,
a conjecture exists \cite{skobelev} that the Feynman rules in the
high magnetic field limit may be directly served by
two-dimensional (one space + one time)  form of electron
propagators. As applied to the Bethe-Salpeter equation, the
dimensional reduction in high magnetic field was considered in
\cite{leinson1}, \cite{ShUs1}, \cite{ShUs}, \cite{leinson2}. In
these references the well-known simultaneous approximation to the
Bethe-Salpeter equation taken in the integral form was exploited,
appropriate for nonrelativistic treatment of the relative motion
of the two charged particles. Once we shall  in the next Section
be interested in the ultrarelativistic regime, we reject from
using this approximation, and find it convenient to deal only with
the differential form of the Bethe-Salpeter equation.

The electron-positron bound state is described by the
Bethe-Salpeter amplitude (wave function)
{\Large$\chi$}$_{\alpha,\beta}(x^\rme,x^\rmp)$ subject to the
fully relativistic equation \cite{schweber}, which in the ladder
approximation in a magnetic field may be written as
\bee\label{equation}[\rmi\widehat{\partial}^\rme-m+e\widehat{A}(x^\rme)]
_{\alpha \beta}
[\rmi\widehat{\partial}^\rmp-m-e\widehat{A}(x^\rmp)]_{\mu\nu}
\begin{tabular}{c}{\Large$\chi$}$_
{\beta\nu}(x^\rme,x^\rmp)$\end{tabular}=\nonumber\\
=-\rmi 8\pi\alpha D_{i
j}(x^\rme-x^\rmp)[\gamma_i]_{\alpha\beta}[\gamma_j]_{\mu\nu}
\begin{tabular}{c}{\Large$\chi$}$_
{\beta\nu}(x^\rme,x^\rmp)$\end{tabular} \eend Here $x^\rme,x^\rmp$
are the electron and positron 4-coordinates,
$D_{ij}(x^{\rme}-x^{\rmp})$ is the photon propagator, and we have
explicitly written the spinor indices
$\alpha,~\beta,~\mu,~\nu=1,2,3,4.$

 We refer to, if needed,
the so called spinor representation of the Dirac $\gamma$-matrices
in the block form\bee\label{gamma}
\gamma_0=\left(\begin{tabular}{cc}0&I\\I&0\end{tabular}\right),\qquad
\gamma_k=\left(\begin{tabular}{cc}0&$-\sigma_k$\\$\sigma_k$&0\end{tabular}
\right),\eend $\sigma_k$ are the Pauli matrices:
 \bee\label{pauli} \sigma_1=\left(\begin{tabular}{cc} 0 &1\\1 &
 0\\
\end{tabular}\right),\qquad \rmi\sigma_2=\left(\begin{tabular}{cc} 0 &1\\-1&
0\\
\end{tabular}\right),\qquad \sigma_3=\left(\begin{tabular}{cc} 1& 0\\0&
-1\\\end{tabular}\right),\eend $k=1,2,3;$~~$m$ is the electron
mass, $e$ the absolute value of its charge $e=2\sqrt{\pi\alpha}$.
The metrics in the Minkowsky space is $\;{\rm diag}\;
g_{ij}=(1,-1,-1,-1)$. The vector potential of the constant and
homogeneous magnetic field $B$, directed along the axis 3 ($
B_3=B,~ B_{1,2}=0$), is chosen in the asymmetric
gauge\bee\label{asym}A_1(x)=-Bx_2,\;A_{0,2,3}(x)=0.\eend With this
choice, the translational invariance along the directions 0,1,3
holds.

Solutions to equation (\ref{equation}) may be represented in the
form
 \bee\label{trans}
 \begin{tabular}{c}{\Large$
 \chi$}$(x^{\rme},x^{\rmp})$\end{tabular}=\nonumber\\\begin{tabular}{c}
 {\Large$
 \eta$}$(x_0^{\rme}-x_0^{\rmp},x_3^{\rme}-x_3^{\rmp},x^{\rme}_
 {1,2},x^{\rmp}_{1,2})$\end{tabular}
 \exp\{\frac\rmi{2}(P_0(x_0^{\rme}+x^{\rmp}_0)-P_3(x_3^{\rme}+x^{\rmp}
 _3))\}, \eend
 where $P_{0,3}$ are the center-of-mass 4-momentum components of
 the longitudinal motion,
 that expresses the translational invariance along the
 longitudinal directions (0,3). We do not find it convenient to be
 using explicitly
 consequences of the magnetic translation invariance
 \cite{gorkov}. (The general representation for the Bethe-Salpeter
 amplitude that incorporates these features may be found in
 \cite{kouzakov}, \cite{kratkie}).
 Denoting the differences $x_0^\rme-x_0^\rmp=t,$
 $x_3^\rme-x_3^\rmp=z$ we obtain the equation
 \bee\label{differential4}\hspace{-2.5cm}
\left[\rmi{\hat{
\partial_\|}}- \frac{\hat{P_\|}}{2}-m+\rmi
{\hat{\partial^\rme_\perp}}-e\gamma_1A_1(x_2^\rme)\right]_{\alpha\beta}
\left[-\rmi{\hat{
\partial_\|}}-\frac{ \hat{P_\|}}{2}-m+\rmi
{\hat{\partial^\rmp_\perp}}+e\gamma_1A_1(x_2^\rmp)\right]_{\mu\nu}\cdot
 \nonumber\\\hspace{-2cm}\cdot
\begin{tabular}{c}{\Large$
 [\eta$}$(t,z,x_\perp^{\rme,\rm p})${\Large ]}$_{\beta\nu}$
 \end{tabular}=-\rmi 8\pi\alpha D_{i
j}(t,z,x^\rme_{1,2}-x^\rmp_{1,2})~[\gamma_i]_{\alpha\beta}~[\gamma_j]
_{\mu\nu}\begin{tabular}{c}{\Large$[
 \eta$}$ (t,z,x_\perp^{\rme,{\rm
p}})${\Large ]}$_{\beta\nu}$\end{tabular}, \eend where
$x_\perp=(x_1,x_2)$,
$-\hat{\partial}_\perp=\gamma_1\partial_1+\gamma_2\partial_2$,
$((\partial_\perp)_i=\partial_i,\; i=1,2)$, $\hat{
\partial_\|}=\partial_t\gamma_0-
\partial_z\gamma_3$, $\hat{P_\|}=P_0\gamma_0-P_3\gamma_3$.

 \subsection{Fourier-Ritus Expansion in eigenfunctions of the transversal motion}
   Expand the dependence of  solution of equation
(\ref{differential4}) on the transversal degrees of freedom into
the series  over the (complete set of) Ritus \cite{ritus} matrix
eigenfunctions\footnote[6]{Our definition of the matrix
eigenfunctions differs from that of  Ref. \cite{ritus} in that the
longitudinal degrees of freedom are not included and the factor
$\exp (\rmi p_1x_1)$ is separated.} $E_h(x_2)$
\bee\label{ex2}\hspace{-2.5cm}\begin{tabular}{c}[{\Large$\eta$}$(t,z,
x_\perp^{\rme,\rmp})]_{\mu\nu}$\end{tabular}= \sum_{h^\rme
h^\rmp}~\rme^{\rmi p_1^\rme x_1^\rme}
[E^\rme_{h^\rme}(x^\rme_{2})]_\mu^{\alpha^\rme}[E^\rmp_{h^\rmp}
(x^\rmp_{2})]_\nu^{\alpha^\rmp}{\rm e}^{\rmi p_1^{\rm p} x_1^{\rm
p}}\begin{tabular}{c}{ [{\Large$\eta$}$_{h^\rme
h^\rmp}(t,z)]^{\alpha^\rme\alpha^\rmp}$}\end{tabular}.\eend Here
{\Large $\eta$}$_{h^\rme h^\rmp}(t,z)$ denote \textit{unknown}
functions that depend on the differences of the longitudinal
variables, while the Ritus matrix functions $\rme^{\rmi
p_1x_1}E_{h}(x_{2})$ depend on the individual coordinates
$x^{\rme,{\rm p}}_{1,2}$ transversal to the field. The Ritus
matrix functions and the  unknown functions
{\Large$\eta$}$_{h^\rme h^\rmp}(t,z)$ are labelled by two pairs
$h^{\rme},h^{\rmp}$ of quantum numbers $h=(k,p_1,)$, each pair
relating to one out of the two particles in a magnetic field. The
Landau quantum number $k$ runs all nonnegative integers
$k=0,1,2,3...$, while $p_1$ is the particle momentum component
along the transversal axis 1. Recall that the potential $A_\mu(x)$
(\ref{asym}) does not depend on $x_1$, so that $p_1$ does
conserve. This quantum number is connected with the orbit center
coordinate ~ $\widetilde{x_2}$~ along the axis 2 \cite{QM}, $p_1=
- \widetilde{x_2}eB$.

The matrix functions $\rme^{\rmi p_1x_1}E^{\rme,\rmp}_{h}(x_{2})$
for transverse motion in the magnetic field (\ref{asym}), relating
in (\ref{ex2}) to electrons (e) and positrons (p), are $4\times 4$
matrices, formed, in the spinor representation, by four
eigen-bispinors of the operator $(\rmi\hat{\partial}_\perp\pm
e\hat{A})^2$ \bee\label{eigen2}(\rmi\hat{\partial}_\perp\pm
e\hat{A})_{\mu\nu}^2 \rme^{\rmi
p_1x_1}[E^{\rme,\rmp}_h(x_2)]^{(\sigma,\gamma)}_\nu=
-2eBk\rme^{\rmi
p_1x_1}[E^{\rme,\rmp}_h(x_2)]^{(\sigma,\gamma)}_\mu,\eend placed,
as columns, side by side \cite{ritus}. Here the upper and lower
signs relate to electron and positron, respectively,
  while $\sigma=\pm 1$ and $\gamma=\pm 1$ are
eigenvalues of the operators
\bee\label{gammafive}\Sigma_3=\left(\begin{tabular}{cc}$\sigma_3$&0\\0&
$\sigma_3$\end{tabular}\right),\qquad
-\rmi\gamma_5=\left(\begin{tabular}{cc}-I&0\\0&I\end{tabular}
\right), \eend diagonal in the spinor representation, to which the
same 4-spinors are eigen-bispinors\footnote {Henceforth, if
superscripts ~e~ or ~p~ are omitted, the corresponding equations
relate both to electrons and positrons in a form-invariant way.}
\bee\label{sg}-\rmi\gamma_5E_h^{(\sigma,\gamma)}=\gamma E_h^
{(\sigma,\gamma)},\quad\Sigma_3E_h^{(\sigma,\gamma)}=\sigma E_h^
{(\sigma,\gamma)}.\quad sg \eend The couple of indices
$\alpha=(\sigma, \gamma)$ is united into one index $\alpha$ in the
expansion (\ref{ex2}), $\alpha=1,2,3,4$ according to the
convention: $(+1,-1)=1,~\; (-1,-1)=2, \;(+1,+1)=3,\; (-1,+1)=4$.
With this convention, the set of 4-spinors
$[E_h(x_{2}]^{(\sigma,\gamma)}_\mu=E_{h}(x_{2})_\mu^{\alpha}$ can
be dealt with as a $4\times 4$ matrix, the united index $\alpha$
spanning a matrix space, where the usual algebra of $\gamma$
-matrices may act. Correspondingly,  in (\ref{ex2}) the unknown
function {\Large$[\eta$}$_{h^\rme h^\rmp}(t,z)${\Large$]$}$^
{\alpha^\rme\alpha^\rmp}$ is a matrix in the same space, and
contracts with the Ritus matrix function.

Following \cite{ritus}, the matrix functions in expansion
(\ref{ex2}) can
 be written in the block form
 as  diagonal matrices
\bee\label{ritus1}
\rme^{\rmi p_1x_1}
E^{\rme,\rmp}_h(x_{2})=\left(\begin{tabular}{cc}$a^{\rme,\rmp}(h;x_{1,2})
$&0\\0&$a^{\rme,\rmp}(h;x_{1,2})$\end{tabular}\right),
\nonumber\\
a^{\rme,\rmp}(h;x_{1,2})=\left(\begin{tabular}{cc}$a^{\rme,\rmp}_{+1}(h;x_{1,2})$&0\\
0&$a^{\rme,\rmp}_{-1}(h;x_{1,2})$\end{tabular}\right).\eend Here
$a^{\rme,\rmp}_\sigma(h;x_{1,2})$ are eigenfunctions of the two
(for each sign $\pm$) operators $[\left(
(\rmi\partial_\perp)_\alpha\pm eA_\alpha\right)^2 \mp\sigma eB]$,
labelled by the two values $\sigma=1,~-1$
\bee\label{eigen1}
[-\left( (\rmi\partial_\perp)_\alpha\pm eA_\alpha\right)^2
\pm\sigma eB]a^{\rme,\rmp}_\sigma(h;x_{1,2})=-2eBk
a^{\rme,\rmp}_\sigma(h;x_{1,2}),\eend Namely, (we omit the
subscript "1" by $p_1$ in what follows)
\bee\label{asigma}a^{\rme,\rmp}_\sigma(h;x_{1,2})=\rme^{\rmi
px_1}U_{k+\frac{\pm\sigma-1}{2}}\left(\sqrt{eB}\left(x_2\pm\frac{p
}{eB}\right)\right),\quad k=0,1,2...,\eend with
\bee\label{parab1}U_n(\xi)=\exp\left\{-\frac{\xi^2}{2}
\right\}(2^nn!\sqrt{\pi})^{-\frac{1}{2}}H_n(\xi)\eend being the
normalized Hermite functions ($H_n(\xi)$ are the Hermite
polynomials). Equations (\ref{eigen1}) are the same as
(\ref{eigen2}) due to the relation
\bee\label{sootn}(\rmi\hat{\partial}_\perp\pm e\hat{A})^2=
-\left((\rmi\partial_\perp)_\alpha\pm eA_\alpha\right)^2 \pm eB
\Sigma_3 \eend and to eq.(\ref{sg}). Besides, the matrix functions
(\ref{ritus1}) are eigenfunctions to the operator
$-\rmi\partial_1$, as commuting with $\Sigma_3$ and $\gamma_5$
(\ref{gammafive}), and with
$(\rmi\hat{\partial}_\perp+e\hat{A})_{\mu\nu}^2$. The
corresponding eigenvalue $p_1$ does not, however, appear in the
r.-h. side of (\ref{eigen1}) due to the well-known degeneracy of
electron spectrum in a constant
 magnetic field.

The orthonormality  relation for the Hermite functions
\bee\label{orth} \int_{-\infty}^\infty
U_n(\xi)U_{n'}(\xi)\rmd\xi=\delta_{nn'}.\eend implies the
orthogonality of the Ritus matrix eigenfunctions in the form
\bee\label{orth2}\sqrt{eB}\int
E^*_{h}(x_{2})_\mu^{\alpha}E_{h^\prime}(x_{2})_\mu^{\alpha^\prime}\rmd
x_{2}= \delta_{kk^\prime}\delta_{\alpha\alpha^\prime}. \eend As a
matter of fact, the matrix functions $E_h(x_2)$ are real, and we
henceforth omit the complex conjugation sign "$^*$".

The matrix functions $\rme^{\rmi px_1}E^{\rme,\rmp}_h(x_{2})$
 (\ref{ritus1}) commute with the longitudinal part\\ $~\pm\rmi\hat
 {\partial}_\|-\hat{P}_\|/2-m$~ of the Dirac
 operator in (\ref{differential4}), owing to the
 commutativity property\bee\label{comm}[E_h(x_2),\gamma_{0,3}]_-=0,
 \eend and
are \cite{ritus}, in a sense, matrix eigenfunctions of the
transversal
 part of Dirac
 operator (not only of its square (\ref{eigen2}))
\bee\label{vazhnoe}(\rmi\hat{\partial}_\perp \pm
e\hat{A})\rme^{\rmi px_1}E^{\rme,\rmp}_h(x_{2})=
\pm\sqrt{2eBk}\;\rme^{\rmi px_1}E^{\rme,\rmp}_h(x_{2})\gamma_1.
\eend The Landau quantum number $k$ appears here as a "universal
eigenvalue" thanks to the mechanism, easy to trace, according to
which the differential operator in the left-hand side of
eq.(\ref{vazhnoe}) acts as a lowering or rising operator on the
functions (\ref{parab1}), whereas the matrix $\sigma_2$, involved
in $\gamma_2$, interchanges the places the functions $U_{k},$
$U_{k-1}$ occupy in the columns. Contrary to relations, which
explicitly include the variable $\sigma$, whose value forms the
number of the corresponding column, relations (\ref{eigen2}),
(\ref{vazhnoe}), (\ref{comm}), and the first relation in
(\ref{sg}) are covariant with respect to passing to other
representation of $\gamma$-matrices, where the matrix $E_h(x_2)$
may become non-diagonal.
\subsection{Equation for the Fourier-Ritus transform of the Bethe-Salpeter
amplitude.}  Now we are in a position to use expansion (\ref{ex2})
in equation (\ref{differential4}). We left multiply it by
$(2\pi)^{-2}eB~\rme^{-\rmi \overline{p}^\rme
x^\rme_1}E^{\rme}_{\overline{h}^{\rme}}(x_2^{\rme})\rme^{-\rmi
\overline{p}^{\rm p}x^{\rm p}_1}E^{\rm p }_{\overline{h}^{\rm p
}}(x_{2}^{\rm p })$, then integrate over $\rmd^2x^{\rm e
}_{1,2}\;\rmd^2x^{\rm p}_{1,2}$. After  using (\ref{vazhnoe}) and
(\ref{comm}), and exploiting the orthonormality relation
(\ref{orth2}) for the summation over the quantum numbers $h^{\rm
e,p}=(k^{\rm e,p}, p_1^{\rm e,p}),$ the following expression
:\bee\label{levaya}\hspace{-3cm}\left[\rmi{\hat{
\partial_\|}}- \frac{\hat{P_\|}}{2}-m+\gamma_1 \sqrt{2eBk^{\rme}}\right]
_{\alpha\alpha^\rme}\left[-\rmi{\hat{
\partial_\|}}- \frac{\hat{P_\|}}{2}-m-\gamma_1 \sqrt{2eBk^{\rm
p}}\right]_{\mu\alpha^\rmp}\begin{tabular}{c}[{\Large$
\eta$}$_{h^{\rme}h^{\rm
p}}(t,z)]^{\alpha^\rme\alpha^\rmp}$\end{tabular}\eend is obtained
for the left-hand side of the Fourier-Ritus-transformed equation
(\ref{differential4}). We omitted the bars over the quantum
numbers.

  Taking the
expression\bee\label{photon1}D_{ij}\left(t,z,x_{1,2}^\rme-x_{1,2}^\rmp)=
  \frac{g_{ij}}{\rmi 4\pi^2}(t^2-z^2-(x_1^\rme-x_1^\rmp)^2-(x_2^\rme-x_2^
  \rmp)^2\right)
  ^{-1}, \eend for the photon propagator in the Feynman gauge, we may then
  write the
  right-hand side of Ritus-transformed
 equation (\ref{differential4})
 as\bee\label{rhs}\hspace{-3cm}\frac\alpha{2\pi^3}
\int\rmd p^\rme~\rmd p^\rmp~ \sum_{k^\rme k^\rmp}~~g_{ij}~\int
[E^{\rme}_{\overline{h}^{\rme}}(x_{2}^{\rme}) \gamma_i
E^e_{h^e}(x^e_{2})]_{\alpha\alpha_\rme}\;
[E^p_{\overline{h}^p}(x^p_{2}) \gamma_jE^{\rm p }_{{h}^{\rm p
}}(x_{2}^{\rm p })]_{\mu\alpha^\rmp}\begin{tabular}{c}[{\Large$
\eta$}$_{h^{\rme}h^{\rm
p}}(t,z)]^{\alpha^\rme\alpha^\rmp}$\end{tabular}\nonumber
\\
\frac{\rme^{\rmi (p^\rme-\overline{p}^\rme )x_1}~\rme^{\rmi
(p^{\rm p}-\overline{p}^{\rm p} )x_1}~\;eB\rmd^2 x_{1,2}^e~\rmd^2
x_{1,2}^p}
{z^2+(x_1^\rme-x_1^\rmp)^2+(x_2^\rme-x_2^\rmp)^2-t^2},\eend
Integrating explicitly the exponentials in (\ref{rhs}) over the
variable $X=(x_1^\rme+x_1^\rmp)/2$, we obtain the following
expression: \bee\label{rhs2}\hspace{-2cm}
\frac\alpha{\pi^2}\int\rmd p~\rmd P_1~\delta(\overline{P}_1-P_1)
\sum_{k^ek^p}~~g_{ij}\int[E^{\rme}_{\overline{h}^{\rme}}(x_{2}^{\rme})
\gamma_i E^e_{h^e}(x^e_{2})]_{\alpha\alpha_\rme}\;
[E^p_{\overline{h}^p}(x^p_{2}) \gamma_jE^{\rm p }_{{h}^{\rm p
}}(x_{2}^{\rm p })]_{\mu\alpha^\rmp}\cdot\nonumber\\
\begin{tabular}{c}[{\Large$ \eta$}$_{h^{\rme}h^{\rm
p}}(t,z)]^{\alpha^\rme\alpha^\rmp}$\end{tabular} \frac{\exp (\rmi
x(\overline{p}-p))\rmd x}
{z^2+x^2+(x_2^\rme-x_2^\rmp)^2-t^2}\;eB\rmd x_2^\rme\rmd x_2^{\rm
p}, \eend where the new integration variables
 $x=x_1^\rme-x_1^\rmp$, $P_1=p^\rme+p^\rmp$, $p=(p^\rme-p^\rmp)/2$ and
 the new definitions
 $\overline{P}_1=\overline{p}^\rme+\overline{p}^\rmp$, $\overline{p}=
 (\overline{p}^\rme-\overline{p}^\rmp)/2$ have been introduced. The pairs of
quantum numbers in (\ref{rhs2}) are\bee\label{pairs2}
\overline{h}^{\rme,\rmp}=(\overline{k}^{\rme,\rmp},
\frac{\overline{P}_1}{2}\pm\overline{p}),\qquad
{h}^{\rme,\rmp}=(k^{\rme,\rmp}, \frac{{P}_1}{2}\pm{p}).\eend Hence
the arguments of the functions (\ref{asigma}) in (\ref{rhs2})
are:\bee\label{arg}\hspace{-3cm}
\sqrt{eB}\left(x_2^\rme+\frac{\overline{P}_1+2\overline{p}}{2eB}\right),
\quad \sqrt{eB}\left(x_2^\rme+\frac{{P}_1+2{p}}{2eB}\right),\quad
\sqrt{eB}\left(x_2^{\rm p}
-\frac{\overline{P}_1-2\overline{p}}{2eB}\right),
\sqrt{eB}\left(x_2^{\rm p} -\frac{{P}_1-2{p}}{2eB}\right),\quad
\quad \nonumber\\\eend successively  as the functions
$E_h(x_{1,2})$ appear in (\ref{rhs2}) from left to right.
 After
fulfilling the integration over $\rmd P_1$ with the use of the
$\delta$-function, introduce the new integration variable
$q=p-\overline{p}$ instead of $p$, and the integration variables
$\overline{x}_2^\rme=x_2^\rme+(\overline{P}_1+2\overline{p})/2eB$,
$\overline{x}_2^\rmp=x_2^\rmp-(\overline{P}_1-2p)/2eB$ instead of
$x_2^\rme$ and $x_2^\rmp$. Then  eq.(\ref{rhs2}) may be written as
\bee\label{rhs3}\hspace{-2cm}\frac\alpha{\pi^2}\int\rmd q~
\sum_{k^\rme k^p}~~g_{ij}~\int
[E^{\rme}_{\overline{h}^{\rme}}(x_{2}^{\rme}) \gamma_i
E^e_{h^e}(x^e_{2})]_{\alpha\alpha_\rme}\;
[E^p_{\overline{h}^p}(x^p_{2}) \gamma_jE^{\rm p }_{{h}^{\rm p
}}(x_{2}^{\rm p })]_{\mu\alpha^\rmp}\begin{tabular}{c}[{\Large$
\eta$}$_{h^{\rme}h^{\rm
p}}(t,z)]^{\alpha^\rme\alpha^\rmp}$\end{tabular}\cdot\nonumber\\
\int\frac{\exp (-\rmi xq)\;\rmd x \;eB\;\rmd\overline{x}^e_2\;\rmd
\overline{x}_{2}^p} {z^2+
x^2+\left(\overline{x}_2^\rme-\overline{x}_2^\rmp-\frac{\overline{P}_1-q}
{eB} \right)^2-t^2}\eend  Now the pairs of quantum numbers in
(\ref{rhs3}) are\bee\label{pairs3}
\overline{h}^{\rme,\rmp}=(\overline{k}^{\rme,\rmp},
\frac{\overline{P}_1}{2}\pm\overline{p}),\qquad
{h}^{\rme,\rmp}=(k^{\rme,\rmp}, \frac{\overline{P}_1}{2}\pm
q\pm\overline{p}).\eend Hence the arguments of the functions
(\ref{asigma}) in (\ref{rhs3}) from left to right are:
\bee\label{arg3}\hspace{-1cm}\sqrt{eB}\overline{x}^\rme_2,\quad
\sqrt{eB}\left(\overline{x}^\rme_2+\frac{q }{eB}\right),\quad
\sqrt{eB} \left(\overline{x}^\rmp_2-\frac{q}{eB}\right),\quad
\left(\sqrt{eB}\overline{x}^\rmp_2\right). \eend

\subsection{Adiabatic approximation.}
  Now we aim at passing to the large magnetic field regime
  in the Bethe-Salpeter equation, with (\ref{levaya}) as the
  left-hand side and (\ref{rhs3}) as the right-hand side.
Define the dimensionless integration  variables $w=x\sqrt{eB}$,
$q'=q/\sqrt{eB}$, $\xi^{\rm e,p}=\overline{x}^{\rm
e,p}_2\sqrt{eB}$ in eq.(\ref{rhs3}). Then it takes the
form\bee\label{rhs4}\hspace{-2cm} \frac\alpha{\pi^2}\int\rmd q'~
\sum_{k^\rme k^\rmp}~~g_{ij}~\int
[E^{\rme}_{\overline{h}^{\rme}}(x_{2}^{\rme}) \gamma_i
E^\rme_{h^\rme}(x^e_{2})]_{\alpha\alpha_\rme}\;
[E^\rmp_{\overline{h}^\rmp}(x^\rmp_{2}) \gamma_jE^{\rm p
}_{{h}^{\rm p }}(x_{2}^{\rm p
})]_{\mu\alpha^\rmp}\begin{tabular}{c}[{\Large$
\eta$}$_{h^{\rme}h^{\rm
p}}(t,z)]^{\alpha^\rme\alpha^\rmp}$\end{tabular}\cdot\nonumber\\
\cdot\nonumber\\\int\frac{\exp (-\rmi wq')\rmd w
\rmd\xi^\rme\rmd\xi^{\rm p}} {z^2+ \frac{w^2}{eB}+\frac 1{eB}
\left(\xi^\rme-\xi^\rmp-\frac{\overline{P}_1}{\sqrt{eB}}-q'
\right)^2-t^2}. \eend  The pairs of quantum numbers in
(\ref{rhs4}) are\bee\label{pairs4}
\overline{h}^{\rme,\rmp}=(\overline{k}^{\rme,\rmp},
\frac{\overline{P}_1}{2}\pm\overline{p}),\qquad
{h}^{\rme,\rmp}=(k^{\rme,\rmp}, \frac{\overline{P}_1}{2}\pm
q^\prime\sqrt{eB}\pm\overline{p}).\eend The arguments of the
functions (\ref{asigma}) in (\ref{rhs4}) from left to right
are:\bee\label{arg4} \xi^\rme,\quad \xi^\rme+q',\quad
\xi^\rmp-q',\quad \xi^\rmp.  \eend

When considering the large field behavior we admit for
completeness that the difference between the centers of orbits
along the axis 2
$\;\widetilde{x}^\rme_2-\widetilde{x}^\rmp_2=-\frac{\overline{P}_1}{eB}$
~ may be kept finite, in other words that the transversal momentum
$\overline{P}_1$ grows linearly with the field. We shall see that
that big transversal momenta do not contradict  dimensional
compactification, but produce an extra regularization of the
light-cone singularity.

In the region, where the 2-interval $(z^2-t^2)^{1/2}$ essentially
exceeds the Larmour radius $L_B=1/\sqrt{eB}$,
\bee\label{domain}z^2-t^2 \gg L_B^2\eend one may neglect the
dependence on the integration variables $w$ and later on $\xi_{\rm
e,p}$ in the denominator. Integration over $w$ produces
$2\pi\delta (q')$, which annihilates the dependence on $q'$ in the
arguments (\ref{arg3}) of the Hermite functions, and they all
equalize.

Let us depict this mechanism in more detail. Fulfill explicitly
the integration over $\rmd w$ in (\ref{rhs4}):\bee\label{explicit}
\hspace{-2.5cm}\int\frac{\exp (-\rmi wq')\;\rmd w}
{z^2-t^2+\frac{w^2}{eB}+ \frac{A^2}{eB}}=
\frac{\sqrt{eB}\pi}{\sqrt{z^2-t^2+\frac{A^2}{eB}}}\left(
\theta(q^\prime)\exp\;
(-q^\prime\sqrt{eB(z^2-t^2)+A^2}~)+\right.\nonumber\\\hspace{5cm}
\left.\theta(-q^\prime)\exp\;
(\;q^\prime\sqrt{eB(z^2-t^2)+A^2}\;)\right),\eend
where\bee\label{A} A^2=
\left(\xi^e-\xi^p-\frac{\overline{P}_1}{\sqrt{eB}}-q' \right)^2
\eend and $\theta(q^\prime)$ is the step function,\bee\label{step}
\theta(q^\prime)=\left\{\begin{tabular}{ccc}1 & when
&$q^\prime>0,$
\\$\frac 1{2}$ & when & $q^\prime=0,$\\0&
when &$ q^\prime<0.$
\end{tabular}\right.\eend Due to the decreasing exponential
in (\ref{parab1}) the variables $\xi^{\rme,\rmp}$ do not exceed
unity in the order of magnitude and can be neglected as compared
to  $\frac{\overline{P}_1}{\sqrt{eB}}$ in (\ref{A}). Unless
$q^\prime$ is  large it may be neglected as compared to the same
term in (\ref{A}), too. Then $A^2=\frac{\overline{P}_1^2}{eB}$,
and after (\ref{explicit}) is substituted in (\ref{rhs4}) and
integrated over $\rmd q^\prime$ the contribution comes only from
the integration within the shrinking region $|q^\prime
|<(eB[z^2-t^2+\frac{\overline{P}_1^2}{(eB)^2}])^{-\frac 1{2}}$.
Then $q^\prime$ can be also neglected in the arguments
(\ref{arg4}). If, contrary to the previous assumption, we admit
that $|q^\prime |$ is of the order of
$\frac{\overline{P}_1}{\sqrt{eB}}\sim\sqrt{eB}$ we see that the
exponentials in (\ref{explicit}) fast decrease with the growth of
the magnetic field as $\exp (-eB(z^2-t^2))$, and therefore such
values of $|q^\prime |$ do not contribute to the integration. If
we admit, last, that $|q^\prime |\gg
|\frac{\overline{P}_1}{\sqrt{eB}}|$, we find that the contribution
$\exp (-|q^\prime|\sqrt{eB(z^2-t^2)+(q^\prime)^2}\;)$ from the
integration over such values is still smaller. Thus, we have
justified the possibility to omit the dependence on $q^\prime$ in
(\ref{A}) and in (\ref{arg4}), and also on $\xi^{\rme,\rmp}$ in
(\ref{A}). Now we can perform the integration over $\rmd q^\prime$
to obtain the following expression for
(\ref{rhs4})\bee\label{rhs5}\hspace{-1.5cm}
\frac{2\alpha\pi^{-1}}{z^2+
\frac{{P}_1^2}{(eB)^2}-t^2}\cdot\nonumber\\\hspace{-1.5cm}\cdot\sum_{k^\rme
k^\rmp}g_{ii}\int[E^{\rme}_{\overline{h}^{\rme}}(x_{2}^{\rme})
\gamma_i
E^\rme_{h^\rme}(x^\rme_{2})]_{\alpha\alpha_\rme}\rmd\xi^\rme\;\int
[E^\rmp_{\overline{h}^\rmp}(x^\rmp_{2}) \gamma_iE^{\rm p
}_{{h}^{\rm p }}(x_{2}^{\rm p
})]_{\mu\alpha^\rmp}\rmd\xi^\rmp\begin{tabular}{c}[{\Large$
\eta$}$_{h^{\rme}h^{\rm
p}}(t,z)]^{\alpha^\rme\alpha^\rmp}.$\end{tabular}\eend It remains
yet to argue that the limit (\ref{rhs5}) is valid also when the
term $\frac{\overline{P}_1}{eB}$ is not kept. In this case we no
longer can disregard $q^\prime$ inside $A^2$ when $q^\prime$ is
less than or of the order of unity. But we can disregard $A^2$ as
compared with $eB(z^2-t^2)$ to make sure that the integration over
$\rmd q^\prime$ is restricted to the  region close to zero
$|q|^\prime <(eB(z^2-t^2))^{-1/2}$ and hence set $q^\prime=0$ in
(\ref{arg4}). The contribution of large $q^\prime$ is small as
before.

 The integration over $\xi_{\rm e,p}$ of the
terms with $i=0,3$ in (\ref{rhs5}) yields the Kroneker deltas
$\delta_{k^\rme\overline{k^\rme}}~\delta_{k^\rmp\overline{k^\rmp}}$
due to the orthonormality (\ref{orth}) of the Hermite functions
thanks to the commutativity (\ref{comm}) of the Ritus matrix
functions (\ref{ritus1}) with $\gamma_0$ and $\gamma_3$. On the
contrary, $\gamma_1,~\gamma_2$ do not commute with (\ref{ritus1}).
This implies the appearance of terms, non-diagonal in Landau
quantum numbers, like $\delta_{k^{\rm e}, \overline{k}^{\rm e}\pm
1}$ and $\delta_{k^{\rm p}, \overline{k}^{\rm p}\pm 1}$, in
(\ref{rhs4}), proportional to ($i=1,2$)
:\bee\label{nondiag1}\hspace{-3cm} T^i_{\overline{k}_\rme\pm
1,\overline{k}^{\rm p}\pm 1}= \sum_{k^\rme
k^\rmp}\int[E^{\rme}_{\overline{h}^{\rme}}(x_{2}^{\rme}) \gamma_i
E^\rme_{h^e}(x^\rme_{2})]_{\alpha\alpha_\rme}\rmd\xi^\rme\;\int
[E^\rmp_{\overline{h}^\rmp}(x^\rmp_{2}) \gamma_iE^{\rm p
}_{{h}^{\rm p }}(x_{2}^{\rm p
})]_{\mu\alpha^\rmp}\rmd\xi^\rmp\begin{tabular}{c}[{\Large$
\eta$}$_{h^{\rme}h^{\rm
p}}(t,z)]^{\alpha^\rme\alpha^\rmp}$\end{tabular}
=\nonumber\\\hspace{-2cm}=\sum_{k^\rme
k^\rmp}\left(\begin{tabular}{cc}0&$-\Delta^{i}
_{\overline{k}^\rme k^\rme}$\\
$\Delta^{i}_{\overline{k}^\rme
k^\rme}$&0\end{tabular}\right)_{\alpha\alpha^\rme}
\left(\begin{tabular}{cc}0&$-
\Delta^{i}_{\overline{k}^\rmp k^\rmp}$\\
$\Delta^{i}_{\overline{k}^\rmp
k^\rmp}$&0\end{tabular}\right)_{\mu\alpha^\rmp}\begin{tabular}{c}[{\Large$
\eta$}$_{h^{\rme}h^{\rm
p}}(t,z)]^{\alpha^\rme\alpha^\rmp}$\end{tabular}. \eend Here
$x^{\rme,\rmp}_2$ are expressed in terms of $\xi$ through the
chain of the changes of variables made above starting from
(\ref{rhs}), so that all the arguments of the Hermite functions
have become equal to $\xi$. Besides,\bee\label{pairs5}
h^{\rme,\rmp}=(k^{\rme,\rmp},\overline{p}^{\rme,\rmp}),\;\quad
\overline{h}^{\rme,\rmp}=(\overline{k}^{\rme,\rmp},\overline{p}
^{\rme,\rmp}),\quad p^\rme+p^\rmp=P_1.\eend
\bee\label{Delta12}\Delta^{i}_{\overline{k}k}=\int
a^\prime(\overline{h},x_2)\;\sigma_{i}\;
a^\prime(h,x_2)\rmd\xi,\qquad i=1,2\eend
\bee\label{delta1}\hspace{-2cm}\Delta^{(1)}_{\overline{k}k}=
\int\left(\begin{tabular}{cc}
0&$a^\prime_{+1}(\overline{h},x_2)a^\prime_{-1}(k,x_2)$\\
$a^\prime_{-1}(\overline{h},x_2)a^\prime_{+1}(h,x_2)$&0\end{tabular}
\right)\rmd\xi= \nonumber\\=
\left(\begin{tabular}{cc}0&$\delta_{\overline{k},\;k-1}$\\
$\delta_{\overline{k},\; k+1}$&0\end{tabular}\right), \eend
\bee\label{delta2}\hspace{-2cm}\Delta^{(2)}_{\overline{k}k}=\rmi
\int\left(\begin{tabular}{cc}
0&$-a^\prime_{+1}(\overline{h},x_2)a^\prime_{-1}(k,x_2)$\\
$a^\prime_{-1}(\overline{h},x_2)a^\prime_{+1}(h,x_2)$&0\end{tabular}
\right)\rmd\xi= \nonumber\\=\rmi
\left(\begin{tabular}{cc}0&$-\delta_{\overline{k},\;k-1}$\\
$\delta_{\overline{k},\; k+1}$&0\end{tabular}\right), \eend The
prime over $a$ indicates that the exponential $\exp (\rmi px_1)$
is dropped from the definitions  (\ref{ritus1}) and
(\ref{asigma}). The non-diagonal Kronecker deltas appeared,
because $a^\prime_{\pm 1}(\overline{h},x_2)$ are multiplied by
$a^\prime_{\mp 1}({h},x_2)$ under the action of the
$\sigma_{1,2}$-blocks in $\gamma_{1,2}$ (\ref{gamma}). In the
final form, the matrices in (\ref{nondiag1}) are
\bee\label{matr}\hspace{-2cm}\left(\begin{tabular}{cc}0&$-\Delta^{i}
_{\overline{k} k}$\\
$\Delta^{i}_{\overline{k} k}$&0\end{tabular}\right)=\frac
1{2}\left(\gamma_1(\pm\delta_{\overline{k},k-1}+\delta_{\overline{k},k+1})+
\rmi\gamma_2(\pm\delta_{\overline{k},k-1}-\delta_{\overline{k},k+1})
\right),\eend with the upper sign relating to $i=1$ and the lower
one to $i=2$. Now equation (\ref{differential4}) acquires the
following form, \bee\label{chain}\hspace{-2cm}\left[\rmi{\hat{
\partial_\|}}- \frac{\hat{P_\|}}2-m+\gamma_1 \sqrt{2eBk^{\rme}}\right]
_{\alpha\alpha^\rme}\left[-\rmi{\hat{
\partial_\|}}- \frac{\hat{P_\|}}2-m-\gamma_1 \sqrt{2eBk^{\rm
p}}\right]_{\mu\alpha^\rmp}\begin{tabular}{c}[{\Large$
\eta$}$_{h^{\rme}h^{\rm
p}}(t,z)]^{\alpha^\rme\alpha^\rmp}$\end{tabular}=\nonumber\\\hspace{-3cm}=
\frac{2\alpha\pi^{-1}}{z^2+
\frac{{P}_1^2}{(eB)^2}-t^2}~\left(\sum_{i=0,3}g_{ii}[\gamma_i
]_{\alpha\alpha^\rme}
[\gamma_i]_{\mu\alpha^\rmp}\begin{tabular}{c}[{\Large$
\eta$}$_{h^{\rme}h^{\rm
p}}(t,z)]^{\alpha^\rme\alpha^\rmp}$\end{tabular}-\sum_{i=1,2}T^{(i)}_{{k}^\rme\pm
1,\;{k}^{\rm p}\pm 1}\right),\quad p_1^\rme+p_1^\rmp=P_1. \eend
The bars over quantum numbers are omitted. This equation is
degenerate with respect to the difference of the electron and
positron momentum components $p=(p^\rme-p^\rmp)/2$ across the
magnetic field, but does depend on its transversal center-of-mass
momentum $P_1=(p^\rme+p^\rmp)$.
 This
dependence is present, however, only for sufficiently large
transverse momenta $P_1$.

At the present step of adiabatic approximation we have come, for
high magnetic field, to the chain of equations (\ref{chain}), in
which the unknown function for a given pair of Landau quantum
numbers $k^\rme,k^\rmp$ is tangled with the same function with the
Landau quantum numbers both shifted by $\pm 1$ (in contrast to the
general case of a moderate magnetic field, where these numbers may
be shifted by all positive and negative integers). To be more
precise, the chain consists of two mutually disentangled
sub-chains. The first one includes all functions with the Landau
quantum numbers $k^\rme,\;k^\rmp$ both even or both odd, and the
second includes their even-odd and odd-even combinations. We
discuss the first sub-chain since it contains the lowest function
with $k^\rme=k^\rmp=0$. We argue now that there exists a solution
to the first sub-chain of equations (\ref{chain}), for which all
{\Large$\eta$}$_{k^\rme ,p_1^\rme;\;k^\rmp ,p_1^\rmp}(t,z)$
disappear if at least one of the quantum numbers $k^\rme,k^\rmp$
is different from zero. Indeed, for $k^\rme=k^\rmp=0$ equation
(\ref{chain}) then reduces to the closed set
\bee\label{closed}[\rmi{\hat{
\partial_\|}}- \frac{\hat{P_\|}}2-m]
_{\alpha\alpha^\rme}[-\rmi{\hat{
\partial_\|}}- \frac{\hat{P_\|}}2-m]_{\mu\alpha^\rmp}
\begin{tabular}{c}[{\Large$
\eta$}$_{0,p_1^{\rme};0,p_1^{\rm
p}}(t,z)]^{\alpha^\rme\alpha^\rmp}$\end{tabular}=
\nonumber\\\hspace{-3cm}= \frac{2\alpha\pi^{-1}}{z^2+
\frac{{P}_1^2}{(eB)^2}-t^2}~\sum_{i=0,3}g_{ii}[\gamma_i]
_{\alpha\alpha^\rme}
[\gamma_i]_{\mu\alpha^\rmp}\begin{tabular}{c}[{\Large$
\eta$}$_{0,p_1^{\rme};0,p_1^{\rm
p}}(t,z)]^{\alpha^\rme\alpha^\rmp}$\end{tabular},\quad
p_1^\rme+p_1^\rmp=P_1.\eend  In writing it we have returned to the
initial designation of the electron and positron transverse
momenta $p^{\rme,\rmp}_1$. Denote for simplicity
{\Large$\eta$}$_{k^\rme k^\rmp}=${\Large$\eta$}$_{k^\rme,p_1^\rme;
k^\rmp,p_1^\rmp}(t,z).$ If  we consider equation (\ref{chain})
with $k^\rme=k^\rmp=1$ for {\Large$\eta$}$_{11}$ we shall have a
nonzero contribution in the right-hand side, proportional to
{\Large$\eta$}$_{00}$ coming from $T^i_{k^\rme-1,k^{\rmp}-1}$,
since the other contributions {\Large$\eta$}$_{11},
${\Large$\eta$}$_{22}$,{\Large$\eta$}$_{20}$,
{\Large$\eta$}$_{02}$ are vanishing according to the assumption.
As the left-hand side of equation (\ref{chain}) now contains a
term, infinitely growing with the magnetic field $B$, it can be
only satisfied with the function {\Large$\eta$}$_{11}$, infinitely
diminishing with $B$ in the domain (\ref{domain}) as
\bee\label{eta11}
\hspace{-2cm}\begin{tabular}{c}[{\Large$ \eta$}$_{11}]^{\alpha\mu}$\end{tabular}=
-\frac {1}{2eB}\frac{\alpha\pi^{-1}}{z^2+
\frac{{P}_1^2}{(eB)^2}-t^2}~\sum_{i=0,3}g_{ii}[\gamma_1\gamma_i]_{\alpha\alpha^\rme}
[\gamma_1\gamma_i]_{\mu\alpha^\rmp}\begin{tabular}{c}[{\Large$
\eta$}$_{00}]^{\alpha^\rme\alpha^\rmp}$\end{tabular},\eend in
accord with the assumption made.  Thus, the assumption that all
Bethe-Salpeter amplitudes with nonzero Landau quantum numbers are
zero in the large-field case is consistent. We state that a
solution to the closed set (\ref{closed}) for {\Large$\eta$}$
_{0,p_1^{\rme};0,p_1^{\rm p}}(t,z)$ with all the other components
equal to zero is a solution to the whole chain (\ref{chain}).

The derivation given in this Subsection realizes formally the
known heuristic argument that, for high magnetic field, the
spacing between Landau levels is very large and hence the
particles taken in the lowest Landau state remain in it.
Effectively, only the longitudinal degree of freedom survives for
large $B$, the space-time reduction taking place. Equation
(\ref{closed}) is a fully relativistic two-dimensional set of
equations with two space-time arguments $t$ and $z$ and two
gamma-matrices $\gamma_0$ and $\gamma_3$ involved. Since, unlike
the previous works \cite{leinson1}, \cite{ShUs}, \cite{leinson2},
neither the famous equal-time Ansatz for the Bethe-Salpeter
amplitude \cite{schweber}, nor any other assumption concerning the
non-relativistic character of the internal motion inside the
positronium atom was made, the equation derived is valid for
arbitrary strong binding. It will be analyzed for the extreme
relativistic case in the next Subsection.

The two-dimensional equation (\ref{closed}) is valid in the
space-like domain (\ref{domain}). It is meaningful provided that
its solution is concentrated in this domain. In non-relativistic
or semi-relativistic consideration it is often accepted that the
wave function is concentrated within the Bohr radius $a_{\rm
0}=(\alpha m)^{-1}=0.5\times 10^{-8}$ cm. It is then estimated
that the corresponding analog of  asymptotic equation
(\ref{closed}) holds true when $a_{0}\gg L_B$, i.e. for the
magnetic fields larger than $\alpha^2m^2/e\sim 2.35\times 10^{9}$
Gauss. This estimate, however, cannot be universal and may be
applicable at the most to the magnetic fields close to the lower
bound  where the value of the Bohr radius can be borrowed from the
theory without the magnetic field. Generally, the question, where
the wave function is concentrated, should be answered \textit {a
posteriori} by inspecting a solution to equation (\ref{closed}).
Therefore, one can state, how large the fields should be in order
that the asymptotic equation (\ref{closed}) might be trusted, no
sooner that its solution is investigated. We shall return to this
point when we deal with the ultra-relativistic situation.

Remind that the transverse total momentum component of the
positronium system is connected with the separation between the
centers of orbits of the electron and positron
$\;P_1/(eB)\;=~\widetilde{x}_2^e-\widetilde{x}_2^p~$ in the
transversal plane, so that the "potential" factor in eq.
(\ref{closed}) may be expressed in the following interesting
form\bee\label{interest}\frac{\alpha}
{(x_0^\rme-x_0^\rmp)^2-(x_3^\rme-x_3^\rmp)^2-(\widetilde{x}_2^\rme-
\widetilde{x}_2^\rmp)^2},\eend(\textit{cf} the corresponding form
of the Coulomb potential in the semi-relativistic treatment of the
Bethe-Salpeter equation in \cite{ShUs1}, \cite{ShUs},
\cite{leinson2}- the difference between the potentials in
\cite{ShUs}, \cite{ShUs1} and \cite{leinson2} lies within the
accuracy of the adiabatic approximation). The appearance of
$P_1^2$ in the potential determines the energy spectrum dependence
upon the momentum of motion of the two-particle system across the
magnetic field like in \cite{ShUs1}, \cite{ShUs}, \cite{leinson2},
\cite{lai}.

We shall need equation (\ref{closed}) in a more convenient form.
First, transcribe it
as\bee\label{closed2}(\rmi\overrightarrow{\hat{
\partial_\|}}- \frac{\hat{P_\|}}2-m)
\begin{tabular}{c}{\Large$
\eta$}$_{0,p_1^{\rme};0,p_1^{\rm
p}}(t,z)$\end{tabular}(-\rmi\overleftarrow{\hat{
\partial_\|}}- \frac{\hat{P_\|}}2-m)^{\rm T}=\nonumber\\=
\frac{2\alpha\pi^{-1}}{z^2+
\frac{{P}_1^2}{(eB)^2}-t^2}~\sum_{i=0,3}g_{ii}\gamma_i
\begin{tabular}{c}{\Large$
\eta$}$_{0,p_1^{\rme};0,p_1^{\rm
p}}(t,z)$\end{tabular}\gamma_i^{\rm T} ,\quad
p_1^\rme+p_1^\rmp=P_1.\eend

Here the superscript T denotes the transposition. With the help of
the relation $\gamma_i^{\rm T}=-C^{-1}\gamma_iC$, with $C$ being
the charge conjugation matrix, $C^2=1$, and the anti-commutation
relation $[\gamma_i,\gamma_5]_+=0$, $\gamma_5^2=-1$, we may write
for a new Bethe-Salpeter amplitude $\Theta(t,z)$, defined
as\bee\label{theta2} \Theta(t,z)=
\begin{tabular}{c}{\Large$
\eta$}$_{0,p_1^{\rme};0,p_1^{\rm
p}}(t,z)$\end{tabular}C\gamma_5,\quad \eend the equation
\bee\label{closed3}(\rmi\overrightarrow{\hat{
\partial_\|}}- \frac{\hat{P_\|}}2-m)
\Theta(t,z)(-\rmi\overleftarrow{\hat{
\partial_\|}}- \frac{\hat{P_\|}}2-m)=\nonumber\\=
\frac{2\alpha\pi^{-1}}{z^2+
\frac{{P}_1^2}{(eB)^2}-t^2}~\sum_{i=0,3}g_{ii}\gamma_i
\Theta(t,z)\gamma_i ,\quad p_1^\rme+p_1^\rmp=P_1.\eend The unknown
function $\Theta$ here is a 4$\times$4 matrix, which contains as a
matter of fact only four independent components. In order to
correspondingly reduce the number of equations in the set
(\ref{closed3}), one should note that the $\gamma$-matrix algebra
in two-dimensional space-time should have only four basic
elements. In accordance with this fact, only the matrices
$\gamma_{0,3}$  are involved  in (\ref{closed3}). Together with
the matrix $\gamma_0\gamma_3$ and the unit matrix $I$ they form
the basis, since $\gamma_{0,3}\cdot\gamma_0\gamma_3=\gamma_{3,0}$,
$\gamma_0^2=-\gamma_3^2=(\gamma_0\gamma_3)^2=1$,
$[\gamma_0,\gamma_3]_+= [\gamma_{0,3},\gamma_0\gamma_3]_+=0$.
Using this algebra and the general representation for the solution
\bee\label{repr}\Theta=aI+b\gamma_0+c\gamma_3+d\gamma_0\gamma_3,\quad
\eend  one readily obtains a closed set of four first-order
differential equations for the four functions $a, b,c,d$ of $t$
and $z$. The same set will be obtained, if one replaces in
eq.(\ref{closed3}) and (\ref{repr}) the 4$\times$4 matrices by the
Pauli matrices (\ref{pauli}), subject to the same algebraic
relations, according, for instance, to the rule:
$\gamma_0\Rightarrow\sigma_3,\;\gamma_3\Rightarrow
\rmi\sigma_2,\;\gamma_0\gamma_3\Rightarrow\sigma_1$. Then equation
(\ref{closed}) becomes a matrix equation
\bee\label{closed4}\hspace{-2cm}(\rmi\overrightarrow{\partial_t}\sigma_3+
\overrightarrow{\partial_z}\sigma_2
-\frac{P_0}2\sigma_3+\frac{P_3}2\rmi\sigma_2-m)
\vartheta(t,z)(-\rmi\overleftarrow{\partial_t}\sigma_3-
\overleftarrow{\partial_z}\sigma_2
-\frac{P_0}2\sigma_3+\frac{P_3}2\rmi\sigma_2-m)=\nonumber\\= \frac
{2\alpha\pi^{-1}}{z^2+ \frac{{P}_1^2-t^2}{(eB)^2}} ~\{\sigma_3
\vartheta(t,z)\sigma_3 +\sigma_2 \vartheta(t,z)\sigma_2\},\quad
p_1^\rme+p_1^\rmp=P_1.\quad \eend for a 2$\times$2 matrix
$\vartheta$ \bee\label{repr2}\vartheta=aI+b\sigma_3+\rmi
c\sigma_2+d\sigma_1.\quad \eend Here $I$ is the 2$\times$2 unit
matrix, and functions $a, b, c, d$ are the same as in
(\ref{repr}).

The following remark is in order. The derivation above relates to
the system of two spinor fields, marked by the superscripts e and
p, with opposite charges $\pm e$ and, generally, different masses.
Although we kept the masses equal above, it is easy to restore
their difference by setting $m=m^\rme$ in the left differential
operator and $m=m^\rmp$ in the right one starting with
eq.(\ref{equation}) throughout. The corresponding Bethe-Salpeter
amplitude {\Large$\eta$} is the translationally invariant part
(\ref{trans}) of the matrix element of the chronological product
of the spinor field operators \bee\label{element1}
\begin{tabular}{c}{\Large$\chi$}$_{\beta\nu}(x^\rme,x^\rmp)$\end{tabular}
=<0|T(\psi_\beta^\rme(x^\rme)\psi_\nu ^ \rmp(x^\rmp))|P>\quad
\eend between the vacuum $<0|$ and the bound state $|P>$. Once we
restrict ourselves to the case where one of the particles,
$\psi^\rme$, is an electron and the other, $\psi^\rmp$, is a
positron, we should take $\psi^\rmp=C\overline{\psi}^\rme$ in
(\ref{element1}) and keep the masses equal. Then the
Bethe-Salpeter amplitude of two arbitrary opposite-charged
fermions $\chi$ and the electron-positron Bethe-Salpeter amplitude
$~\varrho=<0|T(\psi_\beta^\rme(x^\rme)\psi_\nu ^
\rme(x^\rmp))|P>~$ are connected as \;{\Large$\chi$}$=\;\varrho
~C^{\rm T}=-\varrho ~C$. It follows from (\ref{closed4}) that the
translationally-invariant part of the Ritus transform of $\varrho$
obeys the same equation as (\ref{closed3}), but with the sign in
front of the hatted terms in the right-most Dirac operator
reversed, as well as the common sign in the right-hand side. The
subsequent $\gamma_5$-transformation in (\ref{theta2}) is useful,
since it gives the possibility to form the Laplace operator in the
subsequent equations.
\subsection{Including an external electric field}
Let us generalize the two-dimensional Bethe-Salpeter equation
obtained in the presence of a strong magnetic field by including
an external electric field, parallel to it, that is not supposed
to be strong, $E\ll B$. To this end we supplement the potential
(\ref{asym}) in equation (\ref{equation}) by two more nonzero
components\bee\label{electric}A_{0}(x_0,x_3), A_3(x_0,x_3))\neq
0,\quad \quad \eend that carry the electric field - not
necessarily constant - directed along the axis 3. We shall use the
collective notations $A_\parallel=(A_{0}, A_3)$, $x_\|=(x_0,x_3)$,
$\hat{\partial}^{\rme,\rmp}_\|=
{\partial}^{\rme,\rmp}_0\gamma_0-{\partial}^{\rme,\rmp}_3\gamma_3$,
$\hat{A}_\|=A_0\gamma_0-A_3\gamma_3.$ We shall not exploit now a
representation like (\ref{trans}), but deal directly with the
Bethe-Salpeter amplitude {\Large$\chi$}$(x^\rme,x^\rmp)$ as a
function of the electron and positron coordinates, and with its
Fourier-Ritus transform {\Large$\chi$}$_{h^\rme h^\rmp}
(x^\rme_\|,x^\rmp_\|)$ connected with {\Large$\chi$}$
(x_\|^{\rme},x_\|^{\rmp};x_\perp^{\rme},x_\perp^{\rmp})$ in the
same way as (\ref{ex2}). In place of equation
(\ref{differential4}) one should
write\bee\label{differentialE}\hspace{-2.5cm} \left[\rmi{\hat{
\partial_\|}^\rme}- e\hat{A_\|}(x_\|^\rme)-m+\rmi
{\hat{\partial^\rme_\perp}}-e\gamma_1A_1(x_2^\rme)\right]_{\alpha\beta}
\left[\rmi{\hat{
\partial_\|}^\rmp}+e\hat{A}_\|(x^\rmp_\|)-m+\rmi
{\hat{\partial^\rmp_\perp}}+e\gamma_1A_1(x^\rmp_2)\right]_{\mu\nu}\cdot
 \nonumber\\\hspace{-2cm}\cdot
\begin{tabular}{c}{\Large$
 [\chi$}$(x^{\rme,\rmp}_\|,x_\perp^{\rme,\rm p})${\Large ]}$_{\beta\nu}$
 \end{tabular}=-\rmi 8\pi\alpha D_{i
j}(t,z,x^\rme_{1,2}-x^\rmp_{1,2})~[\gamma_i]_{\alpha\beta}~[\gamma_j]
_{\mu\nu}\begin{tabular}{c}{\Large$[
 \chi$}$(x^{\rme,\rmp}_\|,x_\perp^{\rme,\rm p})
${\Large ]}$_{\beta\nu}$\end{tabular},\eend Thanks to the
commutativity (\ref{comm}) the rest of the procedure of the
previous Subsection remains essentially the same, and we come, in
place of (\ref{closed}), to the following two-dimensional
equation\bee\label{closedE}\hspace{-2.5cm} \left[\rmi{\hat{
\partial_\|}^\rme}- e\hat{A_\|}(x_\|^\rme)-m\right]_{\alpha\beta}
\left[\rmi{\hat{
\partial_\|}^\rmp}+e\hat{A}_\|(x^\rmp_\|)-m\right]_{\mu\nu}
\begin{tabular}{c}{\Large$
 [\chi$}$_{0,p_1^\rme;0,p_1^\rmp}(x^{\rme}_\|,x^\rmp_\|)${\Large ]}
 $_{\beta\nu}$
 \end{tabular}=\nonumber\\\hspace{-2cm}=\frac{2\alpha\pi^{-1}}{z^2+
\frac{{P}_1^2}{(eB)^2}-t^2}~\sum_{i=0,3}g_{ii}~[\gamma_i]_{\alpha\beta}~[\gamma_j]
_{\mu\nu}\begin{tabular}{c}{\Large$[
 \chi$}$_{0,p_1^\rme;0,p_1^\rmp}(x^\rme_\|, x^\rmp_\|)
${\Large ]}$_{\beta\nu}$\end{tabular}, \eend for a positronium
atom in a strong magnetic field placed in a moderate electric
field, parallel to the magnetic one. In order to apply this
equation to a system of two different oppositely charged particles
interacting with each other through the photon exchange and placed
into the combination of a strong magnetic and an electric field in
the same direction, say a relativistic hydrogen atom, one should
only distinguish the two masses in the first and second square
brackets in the left-hand side.

\section{Ultra-relativistic regime in a magnetic field}
In the ultra-relativistic limit, where the positronium mass is
completely compensated by the mass defect, $P_0=0$, for the
positronium at rest along the direction of the magnetic field
$P_3=0$, the most general relativistic-covariant form of the
solution (\ref{repr})
is\bee\label{repr3}\Theta=I\Phi+\hat{\partial}_\|\Phi_2+\gamma_0
\gamma_3\Phi_3.\quad \eend The point is that $\gamma_0\gamma_3$ is
invariant under the Lorentz rotations in the plane ($t,z$).
Substituting this into (\ref{closed3}) with $P_0=P_3=0$ we get a
separate equation for the singlet component of (\ref{repr3})
 \bee\label{2D}\left( -\Box_2 + m^2\right)
\Phi(t,z)= \frac{
4\alpha\pi^{-1}\Phi(t,z)}{z^2+\frac{P_1^2}{(eB)^2}-t^2}\quad
 \quad \eend  and the set of equations \bee\label{other}
 \left( \Box_2+ m^2\right)
\Phi_3(t,z)= -\frac{
4\alpha\pi^{-1}\Phi_3(t,z)}{z^2+\frac{P_1^2}{(eB)^2}-t^2},\nonumber\\
(-\Box_2+m^2)\partial_t\Phi_2+2m\rmi\partial_z\Phi_3=0,\nonumber\\
(-\Box_2+m^2)\partial_z\Phi_2+2m\rmi\partial_t\Phi_3=0\quad
 \quad \eend for the other two components. Here $\Box_2=
 -\partial^2/\partial t^2+\partial^2/\partial z^2$ is the Laplace
  operator in two dimensions. Note the "tachyonic" sign in front of
  it in the first equation (\ref{other}).

  Let us differentiate the second equation in (\ref{other}) over $z$
   and the third one over $t$ and subtract the results from each other.
    In this way we get that $\Box_2\Phi_3=0$. This, however, contradicts
 the first equation in (\ref{other}). Therefore, only $\Phi_3= 0$ is possible.
 Then, the two second equations in (\ref{other}) are satisfied,
 provided that $(-\Box_2+m^2)\Phi_2=0$. We shall concentrate in
 equation (\ref{2D}) in what follows.

 The longitudinal momentum along
$x_1$, or the distance between the orbit centers along $x_2$,
plays the role of the effective photon mass and a singular
potential regularizator in equation (\ref{2D}). The lowest state
corresponds to the zero value of the transverse total momentum
$P_1=0$. In this case the equation (\ref{2D}) for the Ritus
transform of the Bethe-Salpeter amplitude finally becomes
\bee\label{2Df}\left( -\Box_2+ m^2\right) \Phi(t,z)= \frac{
4\alpha\Phi(t,z)}{\pi(z^2-t^2)}.\quad \quad \quad \quad \quad
\eend

\subsection{Fall-down onto the center in the Bethe-Salpeter amplitude
for high magnetic field. First hypercritical field.} We are going
now to consider the consequences of the fall-down onto the center
phenomenon present in equation (\ref{2Df}), formally valid for an
infinite magnetic field, and the alterations introduced by its
finiteness.

In the most symmetrical case, when the wave function $\Phi
(x)=\Phi(s)$ does not depend on the hyperbolic angle $\phi$ in the
space-like region of the two-dimensional Minkowsky space,
$t=s\sinh\phi,\; z=s\cosh\phi,\;s=\sqrt{z^2-t^2}$  equation
(\ref{2Df}) becomes the Bessel differential
equation\bee\label{last2} -\frac{\rmd^2\Phi}{\rmd s^2}-\frac
1{s}\frac{\rmd\Phi}{\rmd s}+m^2\Phi=\frac{4\alpha}{\pi
s^2}\Phi.\quad  \eend It follows from the derivation procedure in
the previous Section 3 that this equation is valid within the
interval\bee\label{interval}\frac 1{\sqrt{eB}}\ll s_0\leq
s\leq\infty,\quad\eend where the lower bound $s_0$ depends on the
external magnetic field - it should be larger than the Larmour
radius $L_B=(eB)^{-1/2}$ and tend to zero together with it, as the
magnetic field tends to infinity. The stronger the field, the
ampler the interval of validity, the closer to the origin $s=0$
the interval of validity of this equation extends. If the magnetic
field is not sufficiently strong, the lower bound $s_0$ falls
beyond the region where the solution is mostly concentrated and
the limiting form of the Bethe-Salpeter equation becomes
noneffective, since it only relates to the asymptotic (large $s$)
region, while the rest of the $s$-axis is served by more
complicated initial Bethe-Salpeter equation, not reducible to the
two-dimensional form there. This is how the strength of the
magnetic field participates - note, that the coefficients of
eq.(\ref{last2}) do not contain it.

In treating the  falling to the center below we shall be using
$s_0$ as the lower edge of the normalization box (see the
discussion in Section 2). For doing this it is necessary that
$s_0$ be much smaller than the electron Compton length, the only
dimensional parameter in equation (\ref{last}). In this case the
asymptotic regime of small distances is achieved and nothing
in the region $s<s_0$ beyond  the normalization volume - where the
two-dimensional equations (\ref{closed}), (\ref{closed3}),
(\ref{2D}), (\ref{2Df}) and hence (\ref{last2}) are not valid
- may affect the problem, because  this is left behind the event
horizon.

In alternative to this, we might treat $s_0$ as the cut-off
parameter. In this case we have had to extend equation
(\ref{last}) continuously to the region $0\leq s\leq s_0$,
simultaneously replacing the singularity $s^{-2}$ in it by a model
function of $s$, nonsingular in the origin, say, a constant
$s_0^{-2}$. In this approach the results are dependent on the
choice of the model function which is intended to substitute for
the lack of a treatable equation in that region. Besides, the
limit $s_0\rightarrow 0$ does not exist. The latter fact implies
that the approach should become invalid for sufficiently small
$s_0$, i.e., large $B$. We, nevertheless, shall also test the
consequences of this approach later in this section to make sure
that in our special problem the result is not affected any
essentially.

 The crucial difference of
(\ref{last2}) from equation (\ref{last}), which relates to the
case where
 the magnetic field is absent, is the coefficient
1 in front of the
 first-derivative term  instead of 3 (this coefficient is $D-1$,
where $D$ is the dimension of the Minkowsky space in the radial
part of the Laplacian $\Box_D=s^{-D+1}\partial/\partial
s(s^{D-1}\partial/\partial s)$). The coefficient 4 in front of
$\alpha$ in  (\ref{2Df}) and (\ref{last2}) instead of 8
 in  (\ref{last}) is also of geometric
origin: the relation\bee\label{two} \sum_{i,j=0,3}
g_{ij}\gamma_i\gamma_j=2 \quad \eend was used when we passed from
(\ref{closed}) to (\ref{2D}), whereas relation (\ref{four}) was
exploited to pass from (\ref{schweber}) to (\ref{last}). Solutions
of (\ref{last2}) behave near the singular point $s=0$ like
$s^{\sigma}$, where \bee\label{sigma1}
\sigma=\pm2\sqrt{-\frac\alpha{\pi}}.\eend The fall-down onto the
center \cite{QM} occurs, if $\alpha>\alpha_{\rm cr}= 0$,
\textit{i.e.,} unlike (\ref{alphacr}), for arbitrary small
attraction, the genuine value $\alpha =1/137$ included. With the
substitution ~~$\Phi (s)=\Psi (s) /\sqrt{s}$~~ equation
(\ref{last2}) acquires the standard form of a Schr$\ddot{\rm
o}$dinger equation
\bee\label{last3}\hspace{-2.5cm}-\frac{\rmd^2\Psi(s)}{\rmd
s^2}+\frac{-4\frac\alpha{\pi}-\frac
1{4}}{s^2}\Psi(s)+m^2\Psi(s)=0, \qquad s_0\leq s\leq
 \infty,~\quad s_0\gg(eB)^{-1/2}\quad \eend
 Treating the applicability boundary $s_0$ of this equation as
the lower edge of the normalization box, as  discussed above,
$s_0\ll m^{-1}$,  we impose the standing wave boundary condition
(\ref{stand}) on the solution of (\ref{last3})
\bee\label{mcdonald2}\Psi(s)=\sqrt{s}\;K_\nu(ms),\quad \nu=\rmi
2\sqrt{ \frac{\alpha}{\pi}}\quad \eend that decreases at infinity.

Starting with a certain small value of the argument $ms$, the
McDonald function with  imaginary index (\ref{mcdonald2})
oscillates,
 as $s\rightarrow 0$, passing  the zero value infinitely many
 times.
 Therefore, if $s_0$ is sufficiently small the standing wave
boundary condition (\ref{stand}) can be definitely satisfied.
Keeping to the genuine value of the coupling constant $\alpha
=1/137$\;($\nu=0.096~\rmi$) one may ask: what is the largest
possible value $s_0^{\rm max}$ of $s_0$, for which the boundary
problem (\ref{last3}), (\ref{stand}) can be solved? By demanding,
in accord with the validity condition (\ref{interval}) of equation
(\ref{last2}), (\ref{last3}), that the value of $s_0^{\rm max}$
should exceed the Larmour radius\bee\label{B} s_0^{\rm max}\gg
(eB)^{-1/2}\quad {\rm or}\quad B\gg\frac 1{e\;(s_0^{\rm
max})^2}\qquad \qquad \eend one establishes, how large the
magnetic field should be in order that the boundary problem might
have a solution, in other words, that the point $P_0={\bf P}=0$
might belong to the spectrum of bound states of the Bethe-Salpeter
equation  in its initial form (\ref{equation}).

  One can use the asymptotic form of the McDonald function near zero
to see that the boundary condition (\ref{stand}) is satisfied
provided that \bee\label{spectre}
\left(\frac{ms_0}{2}\right)^{2\nu}=
\frac{\Gamma(1+\nu)}{\Gamma^*(1-\nu)} \qquad  \eend or
\bee\label{spectre2}\nu\ln\frac {ms_0}{2}=\rmi\arg\Gamma(\nu+1)-
\rmi \pi n,\qquad n=0,\pm 1,\pm 2,...\qquad \eend Once $|\nu|$ is
small we may exploit the approximation (\ref{euler}) for the
$\Gamma$-function to
get\bee\label{spectre3}\ln\left(\frac{ms_0}{2}\right)=-\frac n{2}
{\sqrt{\frac{\pi^3}{\alpha}}}-C_{\rm E}, \quad n=1,2,...\quad
\eend We have expelled the non-positive integers $n$ from here for
the same reasons as in Section 2 (see eq.(\ref{spectrum3})). The
maximum value for $s_0$ is provided by $n=1$. The Euler constant
$C_{\rm E}$=0.577 contribution is small as compared to $(1/2)
\sqrt{\pi^3/\alpha}=32.588$.  We finally
get\bee\label{spectre4}\ln\left(\frac{ms_0^{\rm
max}}{2}\right)=-\frac 1{2} {\sqrt{\frac{\pi^3}{\alpha}}}-C_{\rm
E} \nonumber\\\hspace{-2.5cm}{\rm or}\nonumber\\s_0^{\rm
max}=\frac 2{m}\exp\left\{- \frac 1{2}
{\sqrt{\frac{\pi^3}{\alpha}}}-C_{\rm E}\right\}\simeq\frac
2{m}\exp\{-33\}\simeq 10^{-14}\frac 1{m}.\quad \eend This is
fourteen orders of magnitude smaller than the Compton length
$m^{-1}=3.9\cdot10^{-11}$cm and makes about $10^{-25}$cm. Now, in
accord with (\ref{B}), if the magnetic field exceeds the first
hypercritical value of \bee\label{final} B^{(1)}_{\rm
hpcr}=\frac{m^2}{4e}\exp\left\{\frac{\pi^{3/2}}
{\sqrt{\alpha}}+2C_{\rm E}\right\}\simeq 1.6 \times
10^{28}~B_0,\quad \eend
 the positronium ground state
with the center-of-mass 4-momentum equal to zero  appears. Here
$B_0=m^2/e=2.17\times 10^{21}$cm$^{-2}$ is the Schwinger critical
field, or $B_0=m^2c^3/e\hbar= 1.22\times 10^{13}$
 Heaviside-Lorentz units, or $B_0=4.4\times 10^{13}$Gauss ($\alpha=e^2/4\pi\hbar c$, $e=4.8\cdot
 10^{-10}\sqrt{4\pi}$ CGSE). Excited positronium states may also
reach the spectral point $P_\mu=0$, but this occurs for magnetic
fields, tens orders of magnitude larger than (\ref{final}) - to be
found in the same way from (\ref{spectre3}) with $n=2,3...$ The
ultra-relativistic state $P_\mu=0$ has the internal structure of a
confined state, i.e. the one whose wave function behaves as a
standing wave combination of free particles near the lower edge of
the normalization box and decreases as $\exp (-ms)$ at large
distances. The effective "Bohr radius", i.e. the value of $s$ that
provides the maximum to the wave function (\ref{mcdonald2}) makes
$s_{\rm max}=0.17 m^{-1}$ (this fact is established by numerical
analysis). This is certainly much less than the standard Bohr
radius $(e^2m)^{-1}$. Taken at the level of 1/2 of its maximum
value, the wave-function is concentrated within the limits  $0.006
~m^{-1}<s<1.1~ m^{-1}$. But the effective region occupied by the
confined state is still much closer to $s=0$. The point is that
the probability density of the confined state is the wave function
squared \textit{weighted with the measure} $s^{-2}\rmd s$
\textit{singular in the origin \cite{shabad}, \cite{shabad2}} and
is hence concentrated near the edge of the normalization box
$s_0=10^{-25}$cm, and not in the vicinity of the maximum of the
wave function. The electric fields at such distances are about
$10^{43}$ volt/cm. Certainly, there is no evidence that the
standard quantum theory should be valid under such conditions.
This remark gives the freedom of applying the theory in Refs.
\cite{shabad}, \cite{shabad2}.

It is interesting to compare the  value (\ref{final}) with the
analogous value, obtained earlier by the present authors (see
p.393 of Ref.\cite{ShUs}) by extrapolating the nonrelativistic
result concerning the positronium binding energy in a magnetic
field to extreme relativistic region:\bee\label{old} \left.B_{\rm
hpcr}\right|_{\rm
NONRELATIVISTIC}=\frac{\alpha^2m^2}{e}\exp\left\{\frac{2\sqrt{2}}
{\alpha}\right\} =B_0\cdot 10^{164}. \eend Such is the magnetic
field that makes the binding energy of the lowest energy state
equal to (-2m). (This is worth comparing with the magnetic field,
estimated \cite{duncan} as $\alpha^2\exp (2/\alpha) B_0$, that
makes the mass defect of the nonrelativistic hydrogen atom
comparable with the electron rest mass. A more exact
nonrelativistic value for this quantity may be found using the
asymptotic consideration in \cite{karnakov}). We see that the
relativistically enhanced attraction has resulted in a drastically
lower value of the hypercritical magnetic field. Note the
difference in the character of the essential nonanalyticity with
respect to the coupling constant: it is $\exp
(\pi\sqrt{\pi}/\sqrt{\alpha})$ in (\ref{final}) and $\exp
(2\sqrt{2}/{\alpha})$ in (\ref{old}). Another effect of
relativistic enhancement is that within the semi-relativistic
treatment of the Bethe-Salpeter equation \cite{ShUs}, as well as
within the one using the Schr$\rm \ddot{o}$dinger equation
\cite{loudon}, only the lowest level could acquire unlimited
negative energy with the growth of the magnetic field, whereas
according to (\ref{spectre3}) in our fully relativistic treatment
all excited levels with $n>1$ are subjected to the f\al~ and can
reach in turn the point $P_\|=0$.

Let us see now, how the result (\ref{final}) is altered if the
cut-off procedure of Ref.\cite{QM} is used. Consider equation
(\ref{last3}) in the domain $s_0<s<\infty$, but replace it with
another equation\bee\label{psi0}-\frac{\rmd^2\Psi_0(s)}{\rmd
s^2}-\frac{\frac {4\alpha}{\pi}+\frac
1{4}}{s_0^2}\Psi_0(s)+m^2\Psi_0(s)=0\eend in the domain $0<s<s_0$.
The singular potential is replaced by a constant near the origin
in (\ref{psi0}). Demand, in place of (\ref{stand}), that
$\Psi_0(0)=0,\;(\Psi_0'(s_0)/\Psi_0(s_0))=(\Psi'(s_0)/\Psi(s_0))$.
Then the result (\ref{final}) will be modified by the
factor\bee\label{factor}\exp\left\{-\frac
2{\sqrt{\frac{4\alpha}{\pi}+\frac
1{4}}\;\cot\left(\frac{4\alpha}{\pi}+\frac 1{4}\right)-\frac
1{2}}\right\}, \eend which may be taken at the value $\alpha=0$.
Thus, the result (\ref{final}) is only modified by a factor of
$\exp(-4/3)\simeq 0.25$. Generally, the estimate of the limiting
magnetic field (\ref{final}) is practically nonsensitive to the
way of cut-off, in other words to any solution of the initial
equation inside the region $0<s<s_0$, where the magnetic field
does not dominate over the mutual attraction force between the
electron and positron. This fact takes place, because the term
$(\pi^{3/2}\sqrt{\alpha})\simeq 65$, singular in $\alpha$, is
prevailing in (\ref{final}), the details of the behavior of the
wave function close to the origin $s=0$ being not essential
against its background. Although numerically the resulting value
of the crucial magnetic field is affected but very little, we must
bear in mind that the whole cut-off approach is not adequate, as
argued in Section 2, and is burdened by the blind extension of the
standard quantum mechanics to the situation, where electron and
positron are brought together closer than $10^{-14}m^{-1}$ ! The
ultra-relativistic state $P_\mu=0$ arising within this approach is
an ordinary bound state, not the confined state described above.
\subsection{Radiative corrections}
Mass radiative corrections should be taken into account by
inserting the mass operator into the Dirac differential operators
in the l.-h. sides of \BS ~(\ref{equation}) or (\ref{closed}). We
shall estimate now, whether this may affect the above conclusions
concerning the positronium mass compensation by the binding
energy. It was believed that the radiative corrections to the
electron mass are able themselves to annihilate the electron mass
due to the interaction of anomalous magnetic moment with the
external magnetic field. However, this moment is not constant, but
becomes negative for sufficiently strong magnetic field
\cite{baier}. So we are left with the primary result that in the
strong magnetic field the  mass of an electron in Landau ground
state grows with the field $B$ as \cite{jancovici}
\bee\label{mass}\widetilde{m}=m\left(1+\frac\alpha{4\pi}\ln^2\frac{B}{B_0}\right).\eend
 For $B\simeq B^{(1)}_{\rm hpcr}$ the corrected mass  makes $\widetilde{m}= 3.45 m$. This
 implies that the mass annihilation
 due to the f\al ~ requires a field somewhat larger than
 (\ref{final}). To determine its value, substitute $\widetilde{m}$ (\ref{mass})
 for $m$ and $L_B=(eB)^{-1/2}$ for $s_0$ into equation (\ref{spectre3}) with
 $n=1$. The resulting equation for the first
 hypercritical magnetic field, modified by the mass radiative
 corrections, $B_{\rm corr}$\bee\label{corr}\left(1+\frac\alpha{4\pi}
 \ln^2\frac{B_{\rm corr}}{B_0}\right)=4\frac{B_{\rm corr}}{B_0}\exp\left\{\frac 1{2}
{\sqrt{\frac{\pi^3}{\alpha}}}+C_{\rm E}\right\}\eend has the
numerical solution: $B_{\rm corr}= 13~ B^{(1)}_{\rm hpcr}$.

We state that this correction, increasing the first hypercritical
value $B^{(1)}_{\rm hpcr}$ by a little more than one order of
magnitude, is not essential bearing in mind the huge values
(\ref{final}) of the latter.

\subsection{Second hypercritical field.}
The same as in Section 2, we may attribute the  O(1.1)-
symmetrical solution (\ref{mcdonald2}), which is a spinorial
singlet, to the vacuum. Its 2-momentum quantum numbers $P_\|$ are
zero, which means that the vacuum state is constant with respect
to the center-of-mass position $x^\rme+x^\rmp$ of its
constituents. On the contrary, with respect to the space-like
interval $s$ between the vacuum constituents  the wave function
decreases if these are taken apart, as stated above in Subsection
4.1; this means that these constituents are strongly localized.
(The vacuum constituents may be thought of as delocalized in the
"internal coordinate space" obtained by mapping the singular point
of coincidence $s=0$ to the negative infinity - see Refs.
\cite{shabad}, \cite{shabad2} for a detailed explanation of
associated matters).

In this subsection we discuss in a qualitative way the situation
that may take place when the magnetic field exceeds the first
hypercritical value (\ref{final}). The eigenvalues of  \BS
~(\ref{closed}) for the total 2-momentum components $P_{0,3}$ of
the e$^+$-e$^-$ system are now expected to shift into the
space-like region, whereas for $B<B_{\rm hpcr}^{(1)}$ the
center-of-mass 2-momentum of the then real pair was, naturally,
time-like.

With $P_{0,3}\neq 0$  equation (\ref{closed}) becomes more
complicated as compared to the case $P_{0,3}=0$ considered above
in this section. So, decomposition (\ref{repr3}) is no longer
sufficient, but should be supplemented by an extra term
$\hat{P}_\parallel \Phi_4$. The resulting set of equations for
$\Phi$'s does not split now, unlike the set (\ref{2D}),
(\ref{other}) did. Nevertheless, at least for far space-like
$P_\parallel$, $ P_\|^2\ll -4m^2,$ the situation can be modelled
by the same equation as (\ref{last3}), but with the large negative
quantity $~~m^2+P_\|^2/4~~$ substituted for  $m^2$. Then the
McDonald function (\ref{mcdonald2}) is replaced by the Hankel
function containing two oscillating exponentials for large
space-like intervals $s$ \bee\label{hankel}
 \exp\left\{\pm\rmi
s\sqrt{\left|m^2+\frac{P_\|^2}4\right|}\right\}\exp\left\{\rmi
P_\|\frac{x_\rme+x_\rmp}2\right\}.\eend and two oscillating
exponentials \bee\label{sigma3}
s^{\pm2\rmi\sqrt{\frac\alpha{\pi}}}\eend for small ones. If one
passes to the Lorentz frame, where $P_0=0, P_3\neq 0$, and sets
the time arguments in the two-time Bethe-Salpeter amplitude equal
to one another: $x_0^\rme=x_0^\rmp$, one finds that the solution
oscillates along the magnetic field with respect to the relative
coordinate $x_3^\rme-x_3^\rmp$ (mutually free particles) and with
respect to the c.m. coordinate $x_3^\rme+x_3^\rmp$ (vacuum
lattice).

We are now in the kinematical domain called sector IV, or
deconfinement sector in Refs.\cite{shabad}, \cite{shabad2}. In
this sector the constituents are free at large intervals and near
the singular point $s=0$. The wave incoming from infinity is
partially reflected, but partially penetrates to the singular
point, the probability of creation of the delocalized (free)
states is determined by the barrier transmission coefficient
\cite{shabad2}. Such states may exist if one succeeds to satisfy
self-adjoint boundary conditions . These are, for instance,
periodic conditions, to be imposed on the lower and upper
boundaries of the normalization volume, in stead of the standing
wave condition (\ref{stand}), appropriate in sector III.  The
corresponding eigenfunctions are studied in \cite{shabad}. The
possibility to obey them is provided again by the f\al.

Now, the delocalized states in two-dimensional Minkowsky space
 correspond to electron and positron that circle along Larmour
orbits with very small radii in the plane orthogonal to the
magnetic field and simultaneously perform, when the interval
between them is large, a free motion along the magnetic field.
They possess magnetic moments and seem to be capable of screening
the magnetic field. This provides the mechanism that prevents the
classical magnetic field from being larger than the hypothetical
value, second hypercritical field, for which the delocalization
first appears.

The tachyonic character of the vacuum state, i.e. the
space-likeness of its 2-momentum quantum number, does not make a
difficulty. The presence of this quantum number implies the break
down of the  invariance  under the Lorentz transformation along
the magnetic field due to the appearance of the lattice in the
frame $P_0=0$ or of a superluminal wave in arbitrary frame.  This
is not a contradiction, since such a wave appears in response to a
simultaneous increase of the constant magnetic field in the whole
space, which  is already  a noncausal procedure.

\section{Conclusion}
In Section 3 we derived  the fully relativistic two-dimensional
form that the differential Bethe-Salpeter equation for the
electron-positron system takes in the limit of infinite constant
and homogeneous magnetic field imposed on the system. We studied
the f\al~ phenomenon inherent in this equation basing on exactly
relativistic treatment of the relative motion of the electron and
positron. Thanks to this phenomenon, at a certain finite value of
the magnetic field (\ref{final}) called here the \textit{first
hypercritical value}  the positronium level deepens so much that
the rest energy of the system is completely compensated for by the
mass defect. The most symmetrical solution of \BS ~ corresponding
to the center-of-mass momentum equal to zero may be attributed to
the vacuum. In Subsection 4.1 we described the vacuum
restructuring that takes place after the magnetic field exceeds
the first hypercritical value in terms of formation of localized
states of the pair, which are either "confined" or tightly bound -
depending on whether the theory of the f\al ~ in Refs.
\cite{shabad}, \cite{shabad2} is appealed to or not. We estimate
in Subsection 4.2 the modification of the first hypercritical
value of the magnetic field that may be introduced by the mass
corrections to the Dirac field propagator in the strong magnetic
field. In Subsection 4.3 we discuss the \textit{second
hypercritical value} of the magnetic field where a lattice appears
in the vacuum and the latter becomes unstable under the
delocalization of the states of the pair, the delocalized charged
particles on the Larmour orbits being capable of screening the
external field and thus setting a limit to its growth.

The above limiting values are obtained within pure quantum
electrodynamics. Up to now, it was accepted that the vacuum of
this theory is stable with any magnetic field, contrary to
electric field and contrary to non-Abelian gauge field theories
like QCD. In spite of the huge values, expected  to be present,
perhaps, only in superconducting cosmic strings \cite{witten}, the
values obtained may be important as setting the limits of
applicability of QED.

As being due to the special, non-perturbational mechanism
described above, the hypercritical field is determined by the
inverse square root of the fine structure constant elevated to the
exponent. This makes it hundred or so orders of magnitude  smaller
than other known typical values \cite{ritus2} of the magnetic
field that may be expected to lead us beyond  the scope of
coverage of QED owing to the lack of asymptotic freedom. For
instance \cite{shabad3}, the photon becomes a tachyon in the
magnetic field of the order of $B_0\exp (3\pi/\alpha)$.
\section*{Acknowledgements}
One of the present authors (A.E.~Sh.) thanks the Weizmann
Institute of Science where the start to this work was given during
his short stay in Rehovot in April 2004. The work was supported in
part by the Russian
 Foundation for Basic Research
 (project no 05-02-17217) and the President of Russia Programme
  for support of Leading Scientific Schools (LSS-1578.2003.2).

\section*{References}


\end{document}